\documentclass[manuscript]{aastex}

\usepackage{natbib}
\usepackage{color}

\slugcomment{For The Astrophysical Journal}

\shorttitle{Vertical Kink Oscillations in A Large-Scale Plasma Curtain}
\shortauthors{Srivastava and Goossens}

\newcommand{\be}{\begin{equation}}
\newcommand{\ee}{\end{equation}}

\begin{document}


\title{X6.9-class Flare Induced Vertical Kink Oscillations in a Large-Scale Plasma Curtain as Observed by SDO/AIA}


\author{A.K.~Srivastava\altaffilmark{1}}
\affil{$^1$Aryabhatta Research Institute of Observational Sciences (ARIES), Manora Peak, Nainital-263002, India}

\author{M.~Goossens\altaffilmark{2}}
\affil{$^2$Centre for mathematical Plasma Astrophysics, Department of Mathematics,  KU  Leuven,
Celestijnenlaan 200B, 3001, Leuven, Belgium}




\begin{abstract}
We present rare observational evidence of vertical kink oscillations in a laminar and diffused large-scale plasma curtain 
as observed by the Atmospheric Imaging Assembly (AIA) on board the Solar Dynamics Observatory (SDO). The X6.9 class flare in the Active Region 11263 on 09 August 2011, induces a global large-scale disturbance that propagates in a narrow lane above the plasma curtain and creates a 
low density region that appears as a dimming in the observational image data.
This large-scale propagating disturbance acts as a non-periodic driver that interacts asymmetrically and obliquely with the top of the plasma curtain, and triggers the observed oscillations. In the deeper layers of the curtain, we find evidence of vertical kink oscillations with two periods (795 s and 530 s). On the magnetic surface of the curtain where the density is inhomogeneous due to the coronal dimming,  non-decaying vertical oscillations are also observed (period $\approx$ 763-896 s). We infer that the global large-scale disturbance triggers vertical kink oscillations in the deeper layers as well as on the surface of the large-scale plasma curtain. The properties of the excited waves strongly depend on the local plasma and magnetic field conditions. 
\end{abstract}



\section{Introduction}

The major concern of this paper is to understand the conditions for the excitation of vertical kink oscillations 
in a diffused and large-scale laminar plasma curtain.   
Kink MHD waves are exceptional in the sense that they are the only waves that displace the axis of a magnetic tube and the tube as a whole.  
For that reason the standing transverse MHD waves observed in coronal loops are interpreted as standing MHD kink waves. 
Apart from the transverse kink waves, there are claims that torsional Alfv\'en waves have been detected in various
magnetic structures at diverse spatial scales in the solar atmosphere (e.g., Erd\'elyi \& Taroyan 2007; 
Jess et al. 2009; and references cited therein). The observed Alfv\'en waves are  potential candidates for heating the solar corona locally as well as for pursuing MHD seismology of the 
localized atmosphere (Harrison et al. 2002; Dwivedi \& Srivastava 2006; Jess et al. 2009; 
Morton et al. 2011; DeMoortel \& Nakariakov, 2012; Mathioudakis et al. 2013, and references cited there).


Horizontal kink oscillations triggered by flares have been observed in imaging observations of solar loops since the era of 
the Transition Region and Coronal Explorer (TRACE) followed by Hinode, Solar Terrestrial Relationship Observatory (STEREO), 
as well as Solar Dynamics Observatory (SDO) (e.g. Aschwanden et al. 1999; Nakariakov et al. 1999; Van Doorsselaere et al. 2008; Verwichte et al. 2009; Aschwanden \& Schrijver 2011, 
Srivastava et al. 2013, and references cited there). Spectroscopic observations of kink waves are also reported for flaring and non-flaring active region 
loops (O'Shea et al. 2007; Erd\'elyi \& Taroyan 2008, Van Doorsselaere et al. 2008, Andries et al. 2009) using respectively the Coronal Diagnostics Spectrometer (CDS)/SoHO and 
the EUV Imaging Spectrometer (EIS)/Hinode. Observations of kink waves are used to obtain estimates of plasma properties and of  local physical conditions of the solar corona (Nakariakov \& Ofman 2001; Goossens et al. 2002a;  Andries et al. 2005a,b,  
Arregui et al. 2007; Goossens et al. 2008; Andries et al. 2009).  
These  weakly compressible kink waves are also observed in the stellar corona of the $\xi$-Boo (Pandey \& Srivastava 2009, Anfinogentov et al. 2013). 

Vertical kink waves have  been detected in the solar atmosphere since the mid-2000, but only a few cases are known 
so far (e.g., Wang \& Solanki 2004; Li \& Gan, 2006; Aschwanden \& Schrijver 2011; White et al., 2012, and references cited there). White et al. (2012) recently reported flare-induced vertical kink oscillations in an 
eruptive loop in the vicinity of a vertical current sheet. The off-limb coronal loop observed 
by White et al. (2012) is very symmetric and elliptical, and the flare disturbances that triggered the vertical kink wave occurred just below the apex of the loop somewhere near the limb.  Selwa et al. (2011) claim that the excitation of vertical kink oscillations depends on the symmetry of the loop with respect to its interaction with the localized periodic drivers.  
They argue that the absence of symmetry prevents vertical kink modes while horizontal 
kink oscillations can be excited. Selwa et al. (2011) used this argument 
to explain why vertical kink oscillation are so rarely observed in the solar corona.
Aschwanden \& Schrijver (2011) have found that  vertical 
kink oscillations can be triggered by flare-generated fast MHD waves. The 
propagation and interaction of large-scale coronal waves  as well as coronal mass ejections (CMEs) can also trigger 
kink oscillations in the solar loops (e.g., Liu et al. 2011, 2012; Wang et al. 2012).
Several theoretical and numerical investigations have tried  to understand the vertical 
and horizontal polarized kink oscillations of coronal loops in the MHD regime 
(e.g., Gruszecki et al. 2006; Selwa et al. 2007; Mc Laughlin \& Ofman 2008; Selwa \& Ofman 2010, and references cited there).
The onset of the appropriate triggering mechanism and physical conditions to excite the 
vertical kink oscillations are rare in the solar corona 
(Selwa et al. 2011; White et al. 2012). Therefore, there are only few observational
reports in the context of the evolution of vertical kink oscillations in the 
solar atmosphere.
However, sometime limitations in the observational base-line may make the 
identification of these  vertical kink oscillations difficult as they might  appear as horizontal kink 
oscillations due to projection effects (Wang et al. 2008; Aschwanden 2009).

Horizontal and vertical kink oscillations can be a useful tool for
constraining  important properties of the localized corona (e.g., magnetic field)
using MHD seismology (Nakariakov \& Ofman, 2001; Arregui et al. 2007; Van Doorsselaere et al., 2008).
The damping/growth of these waves may also provide clues about the role of dynamic plasma ambient (Ruderman 2011).  
MHD seismology based on the observations of multiple harmonics of kink waves, can lead to  estimates of longitudinal 
density stratification (Andries et al. 2005a,b) and to longitudinal variation and expansion of the magnetic field 
(see e.g., Andries et al., 2009, and references cited there). Resonant absorption may be one of the potential mechanisms
that can explain the observed fast damping of  transverse  standing and propagating kink waves. The transfer of energy from the transverse motions to the azimuthal Alfv\'{e}nic 
motions is essential for the damping by resonant absorption (see e.g., Hollweg and Yang, 1988; Goossens et al. 
1992; Ruderman and Roberts, 2002, Goossens et al., 2000b, 2006, 2009, 2011, 2012). 
The transverse kink motions may undergo rapid dissipationless damping because their 
energy is transferred to the azimuthal Alfv\'{e}nic motions. The eventual damping of the azimuthal Alfv\'{e}nic 
motions is not dissipationless and probably  much slower than the dissipationless damping of the transverse motions.  
This mechanism of dissipationless damping  is universal in the sense that it works for both standing 
and propagating waves (see e.g., Terradas et al. 2010, Verth et al. 2010, Pascoe et al. 2010, 2012). 
A  concern about resonant absorption as a dissipationless damping mechanism of transverse 
(horizontal or vertical) kink waves is the lack of the observation of the azimuthal Alfv\'{e}nic 
motions in the vicinity of the resonant position. Other potential mechanisms for the damping  of the
transverse  waves may be  phase-mixing (Ofman \& Davila 1995; Nakariakov et al. 1997), dissipation of kink and Alfv\'en waves waves by small-scale turbulance (Nakariakov et al. 1999; Van Ballegooijen et al. 2011), and even 
the simple relaxation of the active region field lines.
Therefore, study of the damping and dissipation 
of differently polarized transverse kink oscillations in the magnetized solar atmosphere 
is  of significant importance in the context of the coronal heating.

In the present paper, we outline the rare observational evidence of kink vertical transverse oscillation
in different parts of a large-scale laminar plasma curtain formed by  diffused and unresolved thin 
loop threads in the off-limb corona of the western equator. This is the first detection of collective vertical 
transverse oscillations in a large-scale plasma curtain. The large-scale plasma curtain is very  different from 
individual  coronal loop strands for which  MHD seismology was originally devised. The classic version of MHD seismology almost invariably uses the thin-tube or short wavelength  approximation in cylindrical waveguides (Roberts 1981, Aschwanden 2004)  (cf., review by Andries et al. 2009 and references cited there). 
The detection of vertical oscillations in the present large-scale plasma curtain  may require
new modeling attempts  beyond the 
existing theoretical approach for tubular waves and MHD seismology (Roberts 1981; Nakariakov \& Verwichte 2005).
Indeed the spatial dimensions of this large scale laminar plasma curtain 
are very long and  the inhomogeniety length scales of plasma and magnetic field properties are also far longer 
than those  considered  in the presently existing tube-wave theories (Roberts, 1981; Nakariakov \& Verwichte 2005).
The present paper aims to discuss the detection, possible excitation as well as dissipation 
of the vertical kink oscillations seen  in the plasma curtain.
The X6.9 flare blast loci on 09 August 2011, situated in the northward direction on the solar disk, generate  global disturbances that 
interact with this magnetized plasma medium asymmetrically and obliquely. These global disturbances 
most likely excite vertical kink oscillations in the deeper layer as well as at the outer interface of the 
plasma curtain.  The rest of the paper is structured as follows. In section 2 we describe the observations. In section 3 we 
report the detection of vertical kink oscillations and their implications in the observational base-line of Atmospheric 
Imaging Assembly (AIA) onboard Solar Dynamics Observatory (SDO). 
In section 4 we offer a theoretical interpretation of the observed oscillations. The last section is a 
summary and discussions on the major findings of our paper.

\section{Observations}

A X6.9 class solar flare was observed in the active region AR11263 (N17 W69) near the western equatorial limb on 09 August 2011. 
The flare started at 07:48 UT with a peak intensity at 08:05 UT. The flare ended at 08:08 UT and is classified 
as an impulsive event. In the present paper we have used imaging data showing the flare induced oscillatory 
dynamics of a laminar plasma curtain formed by unresolved, diffused and thin loop threads as observed by 
the Atmospheric Imaging Assembly (AIA) onboard the Solar Dynamics Observatory (SDO). The large-scale plasma 
curtain was situated in 
the off-limb corona at the western equator in the southward direction of the flaring active region (cf., Fig. 1). 
AIA has a maximum resolution of 0.6” per pixel and a cadence of 12 s. AIA provides full disk observations of 
the Sun in three ultra-violet (UV) continua at 1600 \AA~, 1700 \AA~, 4500 \AA~, and seven Extreme Ultraviolet (EUV) 
narrow bands at 171 \AA~, 193 \AA~, 211 \AA~, 94 \AA~, 304 \AA~, 335 \AA~, and 131 \AA~, respectively (Lemen et al. 2012). 
Therefore, it provides observations of multi-temperature, high spatial and temporal resolution plasma 
dynamics in the solar atmosphere. For the current analysis of the flare-induced oscillations of 
the diffused plasma curtain, we have used the data recorded by AIA 171 \AA~ filter. The image sequence 
recorded in 171 \AA~ (T$_{f}$$\sim$10$^{6}$ K) reveal information about the inner coronal plasma. 
The flare generated global disturbances and these global disturbances in turn triggered the observed vertical 
transverse oscillations. We use the SDO/AIA data to capture the signature of these rare oscillations 
in the large-scale plasma structure in the solar corona because of the unique position of the magneto-plasma 
structure almost perpendicular to the line-of-sight (LOS).
To constrain the morphological evolution of large-scale disturbances and creation of the narrow dimming lane 
above the plasma curtain, we use temporal image data in form of 
the difference images of 171 \AA~ (cf., Fig.~2). 
The time-series of SDO/AIA data has been reduced by the SSW cutout service \footnote{http://lmsal.com/get\_aia\_data/}.

\section{Detection of Vertical Kink  Oscillations in the Large-Scale Laminar Plasma Curtain by SDO/AIA}

Fig. 1 shows the images of the active region AR 11263 and its surrounding areas on 09 August 2011. 
SDO/AIA 171 \AA~(left) on 08:07 UT and its difference image (right) show (i) the X6.9
flaring region, (ii) the diffused and laminar plasma curtain formed by the unresolved thin loop threads, 
(iii) a narrow off-limb lane of the propagation of flare generated disturbances above the curtain, and 
(iv) on-disk large-scale global disturbances, respectively by red, blue, green, yellow arrows. 
The difference image of AIA 171 \AA~ on 08:07 with the image of the same FOV on 07:50 UT
shows clearly a dimming in the form of an EUV coronal wave after the maximum onset and 
energy release of the X6.9 flare.  The wavefront propagates as a  global disturbance in 
the on-disk part in South-East of the flaring region. However, the wave disturbance does not 
pass through the plasma curtain directly, and some part of the disturbance only propagates through the narrow 
lane above the plasma curtain. 
For comparison, we display the STEREO-A, EUVI 195 \AA\ image at 08:10 UT 
on 09 August 2011 (Fig.1 bottom-panel; Wuelser et al. 2004). The AR 11263 with its X-class flare energy release lies near the centre in the 
North-East quadrant of the Sun. It is clearly evident that the core off-limb loops in the western side 
of the flaring region partly visible in SDO/AIA field-of-view, are large-scale diffused loops
adjoining the two active regions as shown by black arrow. However,  the large-scale diffused plasma curtain
that lies more to  north-southward  is not visible in the STEREO-A. This may 
be due to its lower  density and emissions in the on-disk view of STEREO-A EUVI 195 \AA\ snapshot. In the 
SDO/AIA field-of-view (FOV), the same plasma curtain is visible off the limb, and lies almost perpendicular to the meridional plane and 
to our line-of-sight (LOS). 
This comparison is to show that the plasma curtain is very
different  from  normal loop systems as it is very diffused at long  spatial-scale  and has a  low density
contrast compared to the ambient atmosphere that made it invisible on solar disk.

Fig. 2 shows the time sequence of the difference images in AIA 171 \AA~channel to examine 
the evolution of the large-scale global disturbances and formation of narrow-dimmed lane 
at the top of plasma curtain. After the
flare peak energy release at 08:05 UT, the on-disk part of the global EIT wave moves as an  
almost semi-circular wavefront in the South- East direction with a  speed of $\approx$750 km s$^{-1}$. 
However, it does not propagate in the form of a symmetric wavefront in the direction of the plasma curtain
as observed under the base-line of SDO/AIA observations. 
Nevertheless,  a narrow dimming region is formed above the plasma curtain that allows the fast moving disturbance 
to propagate above the surface of the plasma curtain (see snapshot on 08:10 UT). 
This disturbance  pushes the curtain from the upward side to the downward direction, and the  
vertical transverse oscillations start after 08:10 UT onwards in the certain part of the plasma curtain (cf., VTO-SDO.mpeg).
It should be noted that the triggered oscillations are perpendicular to the LOS almost along the equatorial 
plane, and thus polarized in form of vertical transverse oscillations. 
It should also be noted that the western core loops connecting the two active regions are visible in
STEREO-A field-of-view, however, more southwardly stretched plasma curtain and higer parts as visible in SDO/AIA of this diffused magnetic structure is 
unfortunately not visible in
STEREO-A most probably due to the less density and emissions. 
Therefore, we only estimate the oscillatory properties of the fully visible 
plasma curtain in SDO/AIA. 
Moreover, the unique position of the plasma curtain in SDO/AIA field-of-view provides us 
an opportunity to detect these oscillations in the deeper layers as well as at the surface of the 
plasma curtain. Therefore, to measure and compare the oscillation properties of the deeper as well
as the surface layers of the plasma curtain, we only use the observations of SDO/AIA 171 \AA\ filter data.
Let us now investigate the properties of flare generated vertical transverse 
oscillations in the large-scale plasma curtain as observed by SDO/AIA.

In Fig. 3, we display the Distance-Time diagram (bottom-left) along a vertical slit (top-left) that
is placed vertically over the plasma curtain in the northward side nearer to the flare blast region. 
The vertical transverse  kink oscillations occur in the deeper layer of the curtain, while the surface of 
the curtain  does not show clear oscillatory signature in a similar way. The Distance-Time (bottom-middle) 
along a parallel vertical slit (top-middle) placed at the  middle of the curtain shows  
vertical oscillations in its deeply rooted  almost same layer, as well as  vertical oscillations of its surface. 
The Distance-Time (bottom-right) along a parallel vertical slit (top- right) placed in southward region on the opposite side
of the curtain near to the flaring region shows oscillations that are only clearly visible at the surface, but that appear 
to be smoothed out from the deeper layers. We also examined the distance-time map along the path traced out perpendicular 
to the plasma curtain and parallel to the polar axis (not shown here). We did not find any signature of propagating kink wave 
trains along the length of the plasma curtain at larger spatial scales at its various heights. This means that only localized vertical kink oscillations 
are triggered in  various parts of the plasma curtain as shown in Figure 3.

In Figs.~4-5, we display the power spectrum analysis of these oscillations carried out using standard wavelet technique (Torrence \& Compo, 1998) and randomization method (Linell Nemec \& Nemec 1985, O'Shea et al., 2001). Let us firstly look at the vertical oscillations that occur in the deeper layer,  which is 30-40 Mm below from the dome-shaped curvilinear outer surface of the plasma curtain. The two panels of Fig.~4 show the oscillations at two different spatial parts of the same deeper plasma layer, which is most likely formed by a bunch of  denser and unresolved loop threads (cf., vertical oscillations of the  same  deeper layer in bottom-left and
bottom-middle panels of Fig. 3). In the first spatial part that lies northward in this layer of the  plasma curtain 
and is closer to the flare blast site, we find  vertical oscillations with a period of $\sim$530 s and a maximum amplitude of about $\pm$4.0 Mm.  
In the second spatial part that lies southward in this layer of the plasma curtain away from the flare blast site in its middle, 
we find  vertical oscillations with a period of $\sim$795 s and a maximum amplitude of about ±5.0 Mm. Let us now turn to the 
outer curvilinear surface of the plasma curtain. This  acts as magnetic interface between the magnetized and denser plasma curtain 
and the outer approximately field-free and less dense region. Initially it does not exhibit clear oscillatory motion  in its northward part. 
However, the oscillations are clearly observed near its middle as well as in the southward off-side from the flare blast region. 
The observed periods of these oscillations are between 764-896 s with  maximum amplitudes of $\pm$5.0-6.0 Mm (cf., Fig.~5).
It should be noted that significant oscillatory power is only evident in all these measurements in the 
intensity wavelet, when the transverse oscillations are switched-on. The transverse oscillations 
are not confined to a single thin flux-tube, but they rather evolve over several layers of the laminar plasma curtain.

\section{Interpretation}\label{SECT:DISS}

We interpret the observed oscillations in the deeper layer of the plasma curtain as vertical kink oscillations. 
The  oscillation with a period  of 795 s might be the fundamental kink oscillation. The reason for this suggestion is that  in most cases known so far it is the fundamental mode that is excited during a flare energy release.  The period of the fundamental mode seems to be rather long compared to the  previously detected periods in  isolated coronal loops. However, it should be noted that the  oscillating part in the middle of the plasma curtain is very long  ($L\sim530 Mm$), and the phase speed of the fast mode kink wave (c$_{k}$) is $2L/P\sim$1332 km s$^{-1}$. The plasma curtain has very low  density contrast ($\rho_{e}/\rho_{o}$$\sim$1.0) compared to the background coronal plasma as it is invisible in the on-disk view in STEREO image (cf., Figure 1). Therefore, the observed kink speed (c$_{k}$) of $\sim$1332 km s$^{-1}$ is almost equal to the internal localized Alfv\'en speed of the plasma curtain $C_{Ao}$. This Alfv\'en speed 
is comparable to the typical inner coronal Alfv\'en speed of 1000 km s$^{-1}$. In the same frame of mind  the oscillation with a period of 530 s could be interpreted as the first longitudinal  overtone. The period ratio is clearly smaller than 2.0 but that might be caused by stratification of density along the  magnetic field  (Andries et al., 2005a; Andries et al. 2005b, Andries et al., 2009;  and references cited there). The fundamental and first overtone are  obvious candidates, but  we have no hard evidence to claim that the observed oscillations indeed correspond to these modes. 
If we consider these two vertical kink oscillations as the  first two longitudinal harmonics then the ratio of the periods can be used to estimate the density scale height  as in Andries et al. (2005a). Here we use the approximation derived by McEwan et al.(2008) for the variation of 
P$_{1}$/2P$_{2}$ as a function of the ratio of half loop length (L/2) and $\Lambda_{c}$ to obtain an estimate for the scale height in the middle part of the plasma curtain where these two oscillations are generated. The length $L$ of this layer along its curved path between two opposite
ends is approximately 530 Mm in the projection of the plane. Therefore, the scale height is estimated as $\Lambda_{c}$$\sim$88 Mm. The scale height is very close to the
typical hydrostatic scale-height of the inner corona at 1.0 MK temperature, i.e., $\sim$80 Mm, when we take into account 
the uncertainity on the estimate of  the length of the oscillating part in the middle of plasma curtain. Therefore, we conjecture that this very large plasma curtain is in the hydrostatic equilibrium conditions. This also favours the fact that at the 
longer spatial scales, the corona can be considered and modelled as a system in  the hydrostatic equilibrium. However,  deviations
from hydrostatic equilibrium are likely to occur in local coronal structures with shorter spatial scales in form of the flux-tubes, e.g., isolated loops, coronal X-ray bright points etc (e.g., McEwan et al. 2006,2008; Andries et al. 2009; Srivastava \& Dwivedi 2010; Kumar et al. 2011; Luna-Cardizo et al. 2012, and references cited there).
The amplitudes  initially increase in time and subsequently the oscillations all of a sudden vanish from different spatial positions in the same deeper layer of the plasma curtain. This behaviour and  dissipative nature are similar to the well-known  fast damping  for horizontal transversal standing waves of the coronal loops (Aschwanden et al., 1999; Nakariakov et al., 1999). 
In the typical solar atmosphere for $\mu$=0.6 and $\gamma$=1.67, the magnetic Reynold number can be approximated as $R_{m}=1.9\times10^{-8}l_{o}V_{o}T_{o}^{3/2}/ln\Lambda$ (Priest, 1982). For the portion of the observed plasma curtain that  as a whole undergoes vertical kink oscillations, the length scale, Alfv\'en velocity, typical temperature for the formation of Fe IX emission are respectively 530 Mm, 1332 km s$^{-1}$, and 1.0 MK. For the typical inner coronal density of 10$^{9}$ cm$^{-3}$ and 
temperature of 1.0 MK, the columb logarithm (log$\Lambda$) is $\sim$19.3 (Priest 1982). Therefore the magnetic Reynold number R$_{m}$ is very high for the observed plasma curtain, which is $\sim$7.0 $\times$10$^{14}$. This is seven order of the magnitude higher than the Reynold number of the environment above a typical photospheric spot, i.e., the coronal loop plasma medium (Priest 1982). This clearly indicates that the collective motion of the  plasma is tightly coupled with the magnetic field lines over the length-scale of the observed plasma curtain.
Therefore, the numerical modelling of the kink oscillations in such plasma curtains need the very high magnetic Reynold number environment. 


The rapid disappearance of the excited vertical kink oscillations might  also be due to a change of angle between the line-of-sight and the plane of the  polarization. However, actual  damping mechanisms may cause the  oscillations to disappear. We will discuss the 
most plausible candidates for damping and  dissipation  in the forthcoming sections.

Another possibility is the detection of two independent kink vertical oscillations in  two different parts of the same deeper
layer of the plasma curtain. The origin of these independent vertical kink oscillations can depend on the energy of the driver,
the nature of the interaction with the localized portion of the curtain having specific plasma and magnetic field properties. It is clear
that the  northward deeper layer, which is near to the flare energy release site, supports comparatively high frequency (low period; 530 s) 
vertical oscillations. This is the location that may serve as the interaction region with a more energetic global wavefront that originated near the
flare blast locus. When we examine the same deeper layer situated southward from the flare blast locus, we find  vertical
oscillations with comparatively low frequency (or long period; 796 s). This is the location that
may serve as the  interaction region with a comparatively less energetic front of the global disturbance that already  lost 
part of its energy on its way in the solar corona. In the far-southward situated region of the 
same deeper layer of the curtain,   collective vertical oscillations cannot be clearly identified. This region  is 
the remotely positioned interaction region where the global disturbance  fails to pump sufficient energy 
to trigger oscillations. We conjecture that the global disturbance  transfers its energy
locally to  different parts of the plasma curtain depending upon the local plasma and magnetic field conditions
there. Hence, the global disturbance excites  different vertical oscillatory modes in different regions (Ballai et al., 2005). The study of this  interaction is outside of the scope of this paper. These alternative interpretations also seem likely 
as we could not get any evidence of running kink wave trains across the major axis of the 
large-scale curtain drawn parallel to the solar North-South axis as well as meridian plane. We only have  evidence of localized vertical oscillations in various parts of the curtain as shown in Fig.~3.

We do not aim here to model  the presently observed large-scale system where the oscillations are excited.  But we would like to point out what we understand with the term  "Laminar Plasma Curtain"  system. It cannot be viewed 
as a single magneto-plasma system since  that would violate the equilibrium conditions. On the contrary, as we stated 
in the beginning, the system  is made-up of many thin un-resolved semi-circular diffused loops
and the plasma filling factor is at
larger  spatial scale than the width of a single average fluxtube. The  whole structure resembles  a large-scale curtain, and the word 'laminar' refers to this  structuring in the form  of unresolved fluxtubes.
The observed kink motions are the collective antisymmetric transverse motions of magnetic field and associated plasma. However,  these
perturbations are distributed over a comparatively longer spatial scale, i.e., in more wide regions of the
plasma curtain.

Let us focus on the excitation of the vertical transverse waves of the present investigation.   
The X6.9 class flare in the Active Region 11263 on 09 August 2011 induces a global large-scale disturbance that propagates in a narrow lane above the plasma curtain and creates a dimming region with lower density. This large-scale propagating disturbance 
acts as a non-periodic driver that interacts asymmetrically and obliquely with the plasma curtain, and triggers waves. This differs from the 
observations by White et al. (2012).  White et al. (2012) observed vertical kink oscillations in a flaring loop, with a 
period of 5.0 min. They conjectured that the oscillation is driven by the periodic reconnection and formation of a  post-flare hot 
loop on a scale of a 5.0 min periodicity, and  is not due to  flare-generated blast waves. Here we observe similar 
oscillations in a large-scale and diffused plasma curtain triggered by flare generated global disturbances. Moreover, the observed 
periods of 795 s and 530 s are quite long compared to the 5.0 min scale. Although, the curtain is formed by laminar arrangements of 
thin loop threads we unfortunately, can not resolve them in the present observational base-line. The oscillations are triggered 
unambiguously in that part of the curtain where the plasma layer is slightly denser.  This observational situation should also be confronted with  the numerical findings of Selwa et al. (2011).  Selwa et al. (2011)  theorized that the amplitude of vertical kink oscillations 
in a dipolar coronal loop is significantly amplified in comparison to that of horizontal kink oscillations for a pulse driver 
that is located symmetrically below the loop.  In inclined loops they could not identify vertical kink oscillations. They used 
this scenario to explain the scarsity of vertical kink oscillations in the corona. 
Our observations of vertical kink oscillations show a scenario  that  is completely different from their findings
since here the vertical kink oscillations 
are excited by a  non-periodic and non-symmetric driver, i.e., flare generated global coronal waves. 

Let us now turn to the observations of oscillations on the surface of the plasma curtain. There we 
also observe a vertical kink oscillation, now with a period  between 763 and 896 s.  This oscillation is
uniformly distributed near the middle and southward side of the plasma curtain surface. However, it appears 
to be absent from the surface of the plasma curtain in the northward side that is nearer to the flare site. 
The amplitude of this oscillation does not show any signature of decay in the present available observational 
baseline. 


The curtain lies  almost in the part of the equator near the  west-limb, so that  the density can be taken to be a smooth function of 
the radial outward  distance inside it. In the deeper layers of the plasma curtain the density is almost constant and 
the kink waves have predominantly vertical transverse motions. Above the northward surface of curtain, the dimming is 
comparatively weak  (cf., snapshot 08:10 UT in Fig. 2) and the spatial  changes in density are smaller inside and outside of the surface. The dimmed region is clearly present near the middle of the surface as well as at the surface in 
the southward direction. This implies a clear spatial change in density and local Alfv\'{e}n speed
at different spatial locations (northward, middle and southward) above the surface of the plasma curtain in the 
corona assuming that the  background magnetic field  is constant.  
This 
spatial variation of density and  local Alfv\'{e}n speed increases the relative importance of the azimuthal component of the  
displacement with respect to the radial component at different locations at the surface of plasma curtain.
The vertical kink oscillations 
that  grow at the surface of the plasma curtain during the coronal dimming, have 
velocity components perpendicular to our LOS.

These surface oscillations do not decay while the oscillations  in the deeper layers decay
in the observational base-line of SDO/AIA. The reason may be that the continuous dimming 
due to the passage of fast global disturbances above the curtain creates a low density
region. A variation in density in the radial direction can cause damping of standing and propagating 
waves due to resonant absorption (cf., Goossens et al., 2000b, 2006, 2009, 2011, 2012; Terradas et al. 2010).
Density variation along the loop can impede the damping due to resonant absorption or even cause amplification 
(Soler et al., 2011). Time dependent variation of the background density counteracts the damping due to resonant absorption 
of standing waves and even cause amplification (Ruderman, 2011).
The cooling and variation of the background plasma properties may also 
cause the damping of the loop kink oscillations (Morton \& Erd\'elyi 2009,2010a, Morton et al. 2010b, and 
references cited there). However, this scenario is not in support for the present observations
as we do not get any signature of cooling of the plasma curtain.
Resonant absorption may cause a transfer of
energy of the vertical oscillations in the deeper layers towards the outer surface and cause
their dissipation. The density decrement above the surface due to dimming may counter-act 
this dissipation by resonant absorption and the oscillations may not be dissipated as observed by SDO/AIA. Similar
conditions have been observed by Wang et al. (2012) as a coronal dimming 
above a loop due to CME caused the amplification of kink oscillations. In the present case,
the surface oscillations are not amplified, but remain unchanged and do not undergo dissipation.

\section{Summary}\label{SECT:DISS}

Let us recall the  major concern  of this paper is to understand conditions for the excitation of vertical kink waves 
in a large-scale and diffused plasma curtain. 
It is clear that the kink waves that are observed in the present study are excited by a large scale propagating disturbance that acts as a non-periodic driver and interacts asymmetrically and obliquely with the plasma curtain. This way of excitation differs from that 
active in the observations by White et al. (2012) where the vertical kink waves are excited by a periodic and symmetric driver. 
There is nothing wrong with vertical kink oscillations excited in two different ways. However, our observations contradict
the conjecture by Selwa et al. (2011) that  the excitation of vertical kink oscillations requires a driver that is located symmetrically below the loop.  Further numerical modeling is needed here  keeping the present observations in mind.

The major challenge opened by this first observational findings of  vertical kink oscillations
in the large-scale plasma curtain, is to try and understand  how these waves can be excited 
in such systems beyond the thin tube approximation. If  different vertical oscillations in different 
spatial locations of the plasma curtain are  generated by one-to-one  interactions with  global wave energetic fronts, 
then the evolution of these waves and the transfer of energy  depend on the structure of magnetic field and the composition  
of the medium in which the waves are generated and propagate. Hence MHD models of kink waves  in 
large-scale plasma structures as observed in our paper need to be developed. 
The high magnetic Reynold number plasma environment of the curtain that is almost maintained 
in hydrostatic equilibrium at larger spatial scale is set in the vertical kink mode oscillations, which 
are non-decaying at the surface while quickly decaying on its deeper layers. The non-decaying and decaying kink 
oscillations have been recently observed in coronal loops that are driven respectively by the low amplitude continous and harmonic driver, and 
high amplitude impulsive driver (Nistic{\`o} et al. 2013). They have reported that 
impulsive excitation generates systematically the transverse perturbations in all the loops.
We also do observe the vertical oscillations over larger spatial scales that are generated 
by the impulsive exciter, however, in our case both the decaying and decayless oscillations are seen to be associated with the same impulsive driver. Therefore, the 
opposite nature of the surface oscillations as well as the oscillations 
in the deeper layer of the plasma curtain, indicate about the creation 
of some resonant layers. It is also likely that the resonant absorption is at work in decaying 
the vertical oscillations of the deeper layers of curtain that are associated 
with the impulsive driver. However, the dimming above its surface
most-likely counter act on it and keeping the surface oscillations un-damped. The detailed theoretical
investigations of the excitation and dissipation of such vertical kink oscillations in the
large-scale plasma curtain is the subject of our future work, and outside the scope of this 
paper.

In the present paper, in conclusion, we report a very likely physical scenario for the 
unique observations of the vertical transverse waves in a large-scale plasma curtain.  
However, more study should be made further using space borne observations and stringent 
modeling to shed new light  on the physics of this coronal dynamics.

\section{Acknowledgments}
We thank referee for his valuable suggestions that improved the manuscript considerably.
We acknowledge the use of the SDO/AIA observations for this study. The data are provided by the  courtesy of NASA/SDO, LMSAL, and the AIA, EVE, and HMI science teams. AKS also thanks Shobhna Srivastava for her support and encouragement during this work. MG acknowledges support by the University of Leuven grant GOA/2009-009 and also partial support by the Interuniversity Attraction Poles Programme initiated by the Belgian Science Policy Office (IAP P7/08 CHARM).
AKS acknowledges the support of DST-RFBR (INT/RFBR/P-117) and Indo-Austrian (INT/AUA/BMWF/P-18/2013) project  funds during the present research..

{}

\clearpage

\begin{figure*}
\centering
\mbox{
\includegraphics[scale=0.53, angle=0]{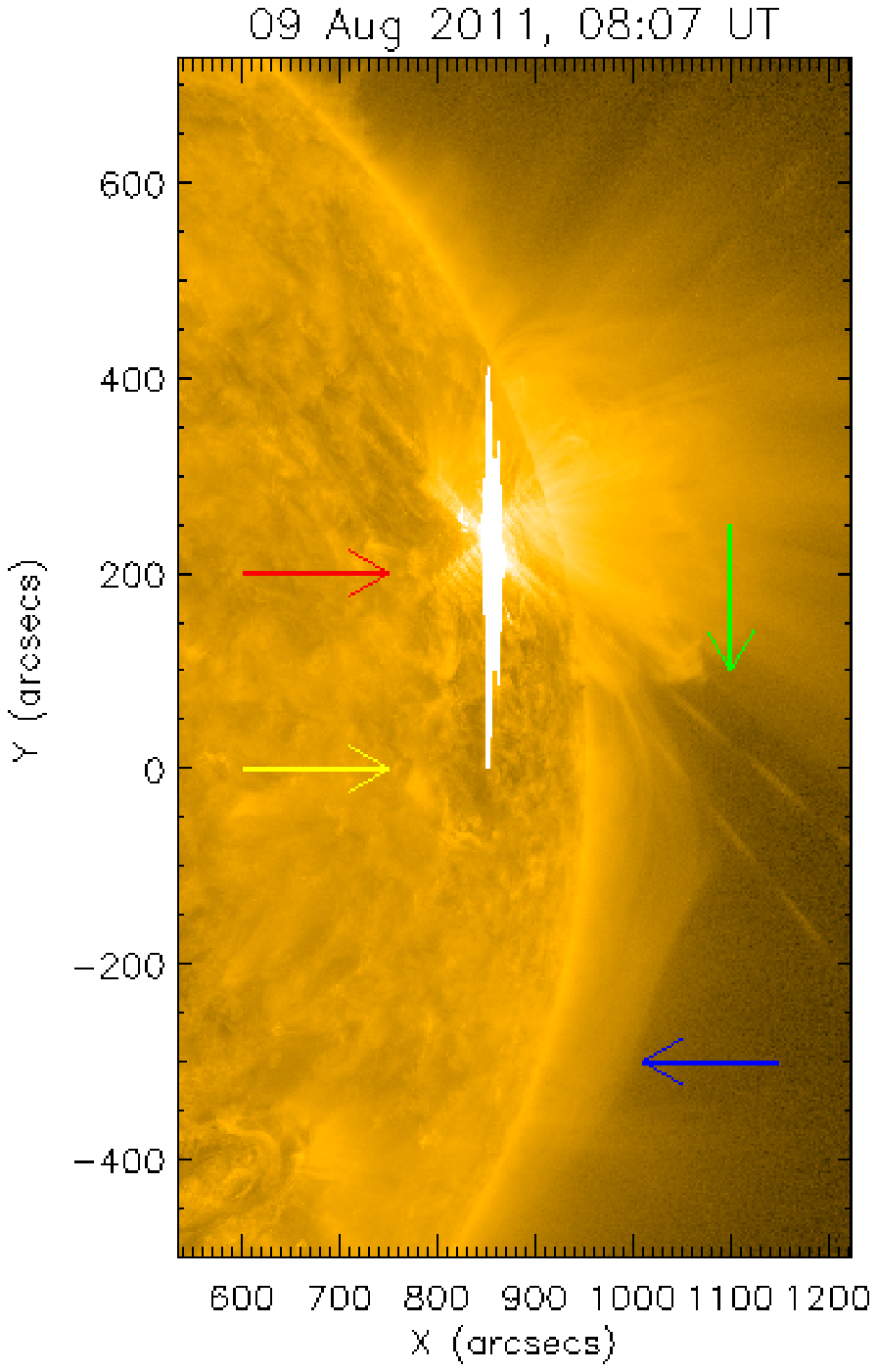}
\hspace{-4.5cm}
\includegraphics[scale=0.53, angle=0]{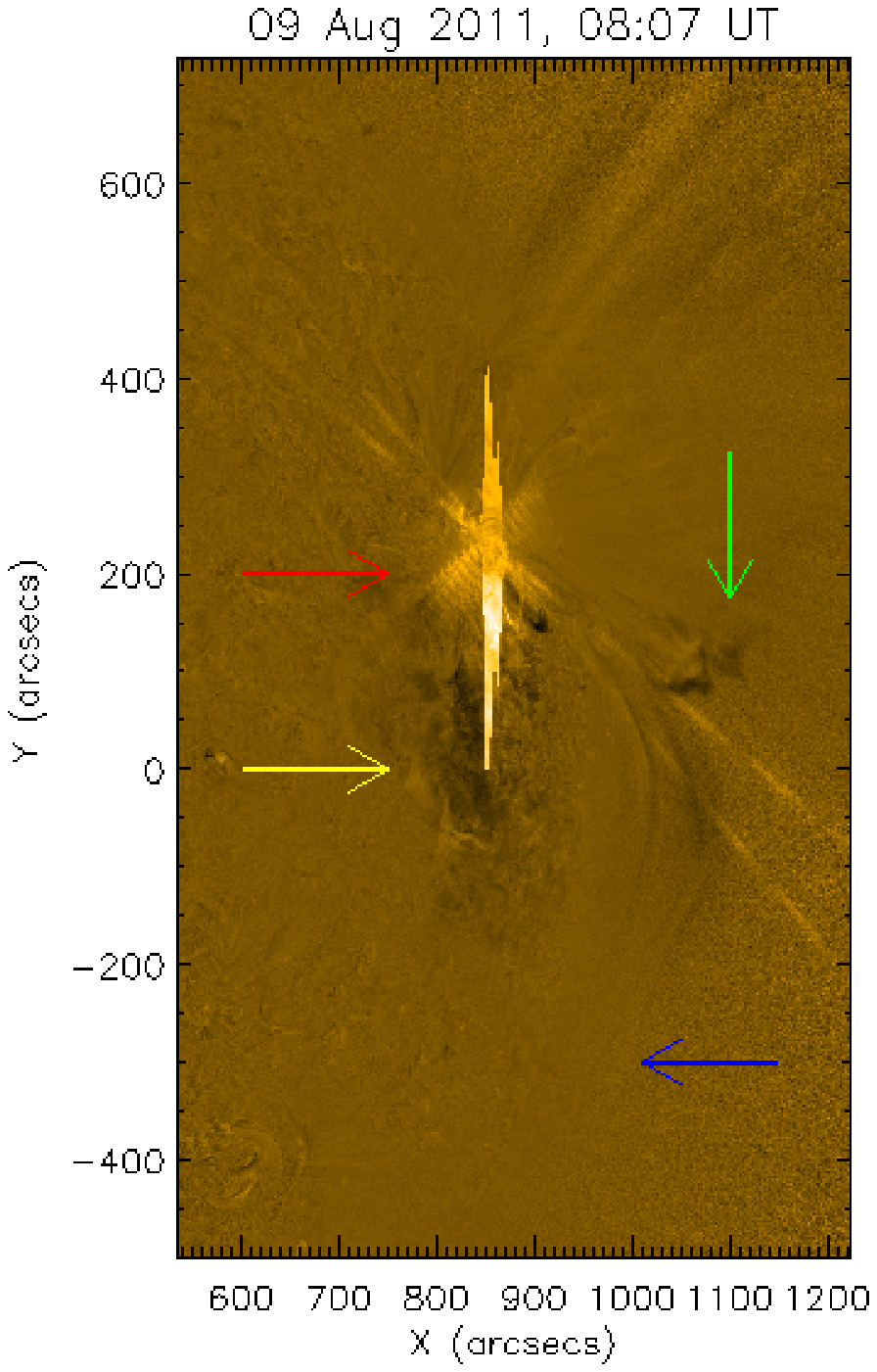}}
\centering
\hspace{+4.0cm}
\includegraphics[scale=0.53, angle=90]{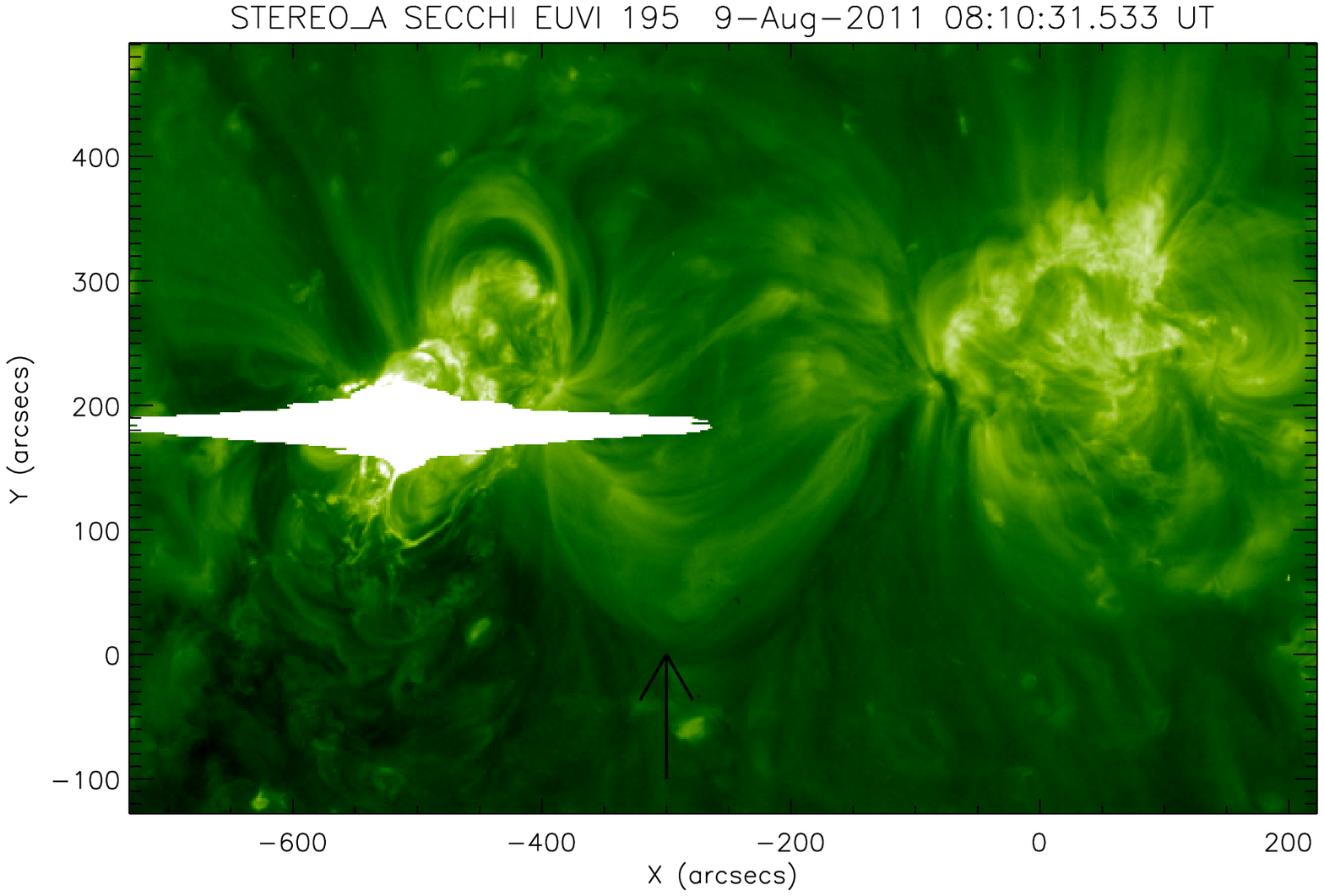}
\caption{\small
SDO/AIA 171 \AA~(top-left) and its difference image (top-right), which display the X6.9 flaring region,
diffused plasma curtain formed by the unresolved thin loop threads, a narrow off-limb lane of 
the propagation of flare generated disturbances, and on-disk large-scale global disturbances,
respetively by red, blue, green, yellow arrows. The bottom panel shows the STEREO-A
EUVI 195 \AA\ image showing the flaring region and western side core loops (indicated by black arrow)
connecting two active regions. However, the less denser and more southwardly stretched 
diffused plasma curtain is not visible.    
}
\label{fig:JET-PULSE_1}
\end{figure*}
%
\clearpage

\begin{figure*}
\centering
\centering
\mbox{
\includegraphics[scale=0.50, angle=0]{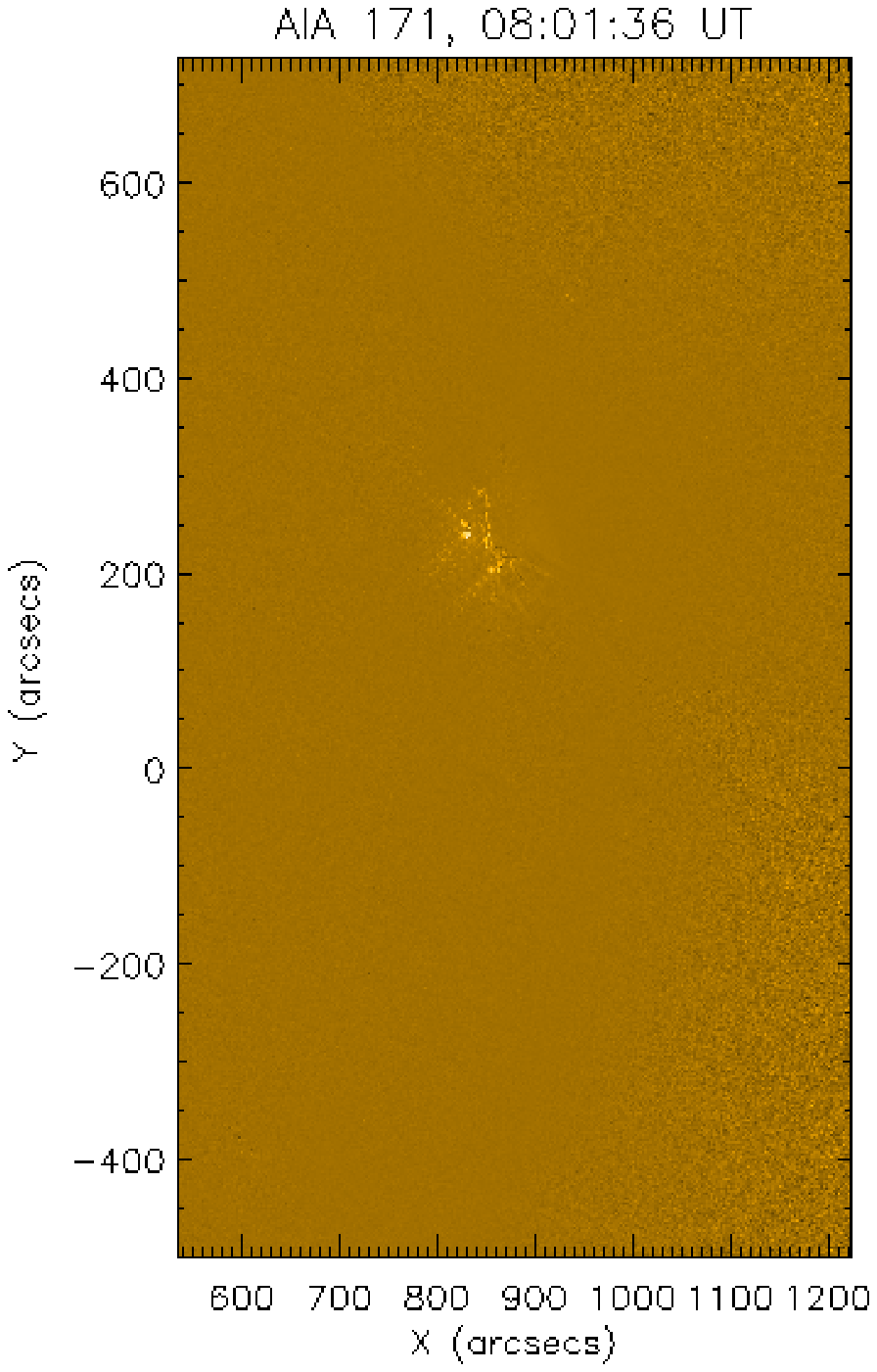}
\includegraphics[scale=0.50, angle=0]{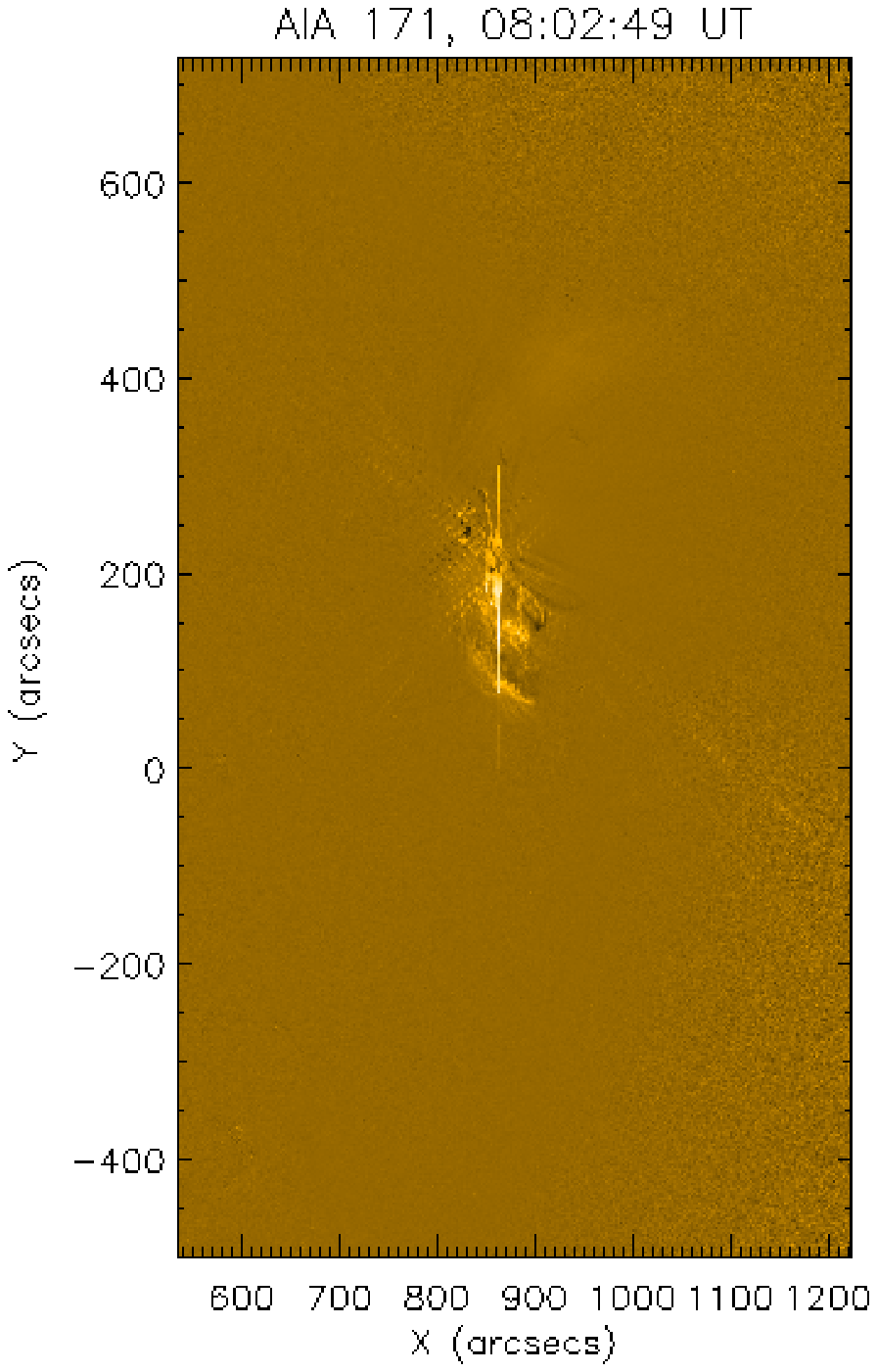}
\includegraphics[scale=0.50, angle=0]{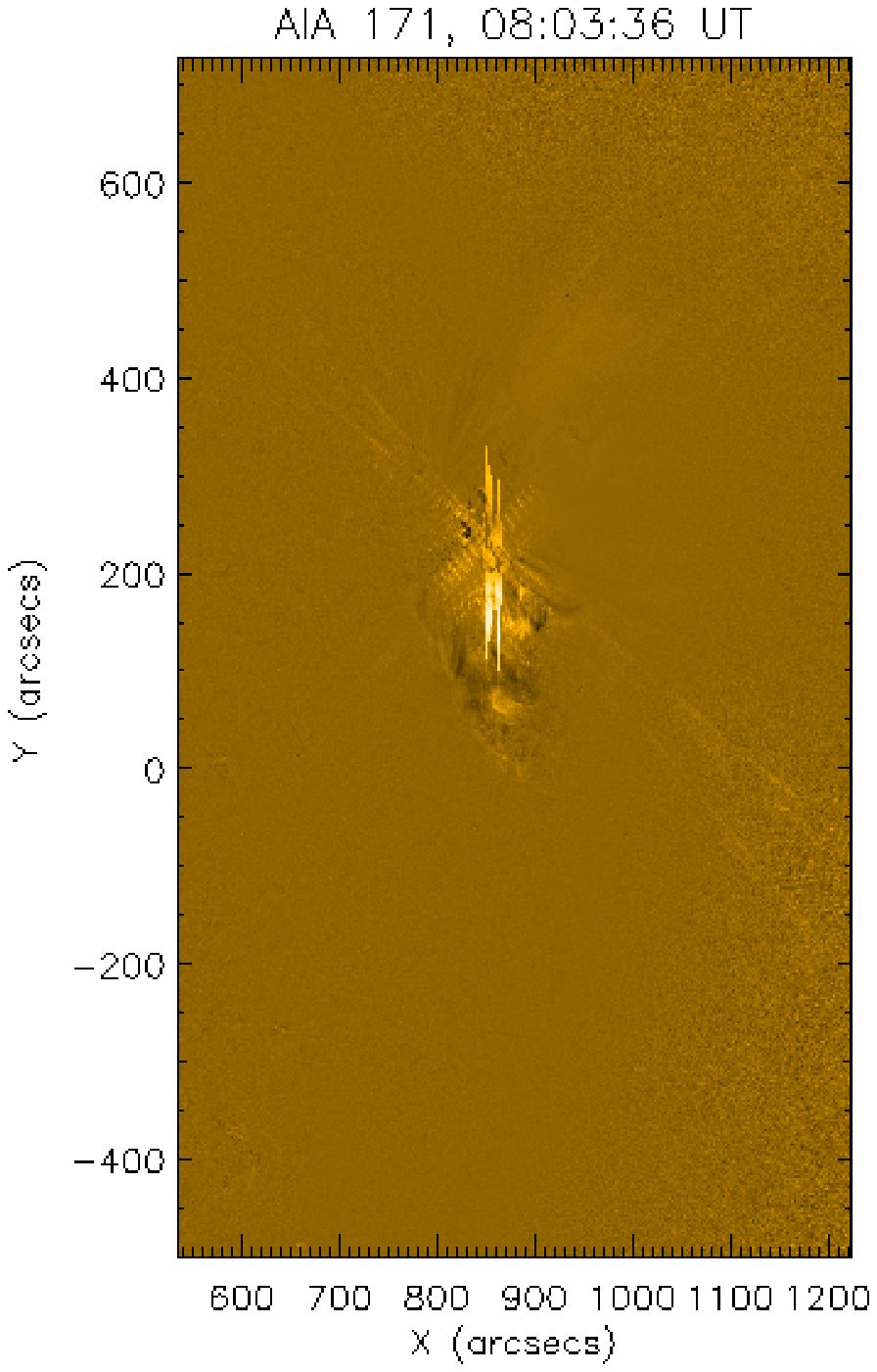}
}

\centering
\mbox{
\includegraphics[scale=0.50, angle=0]{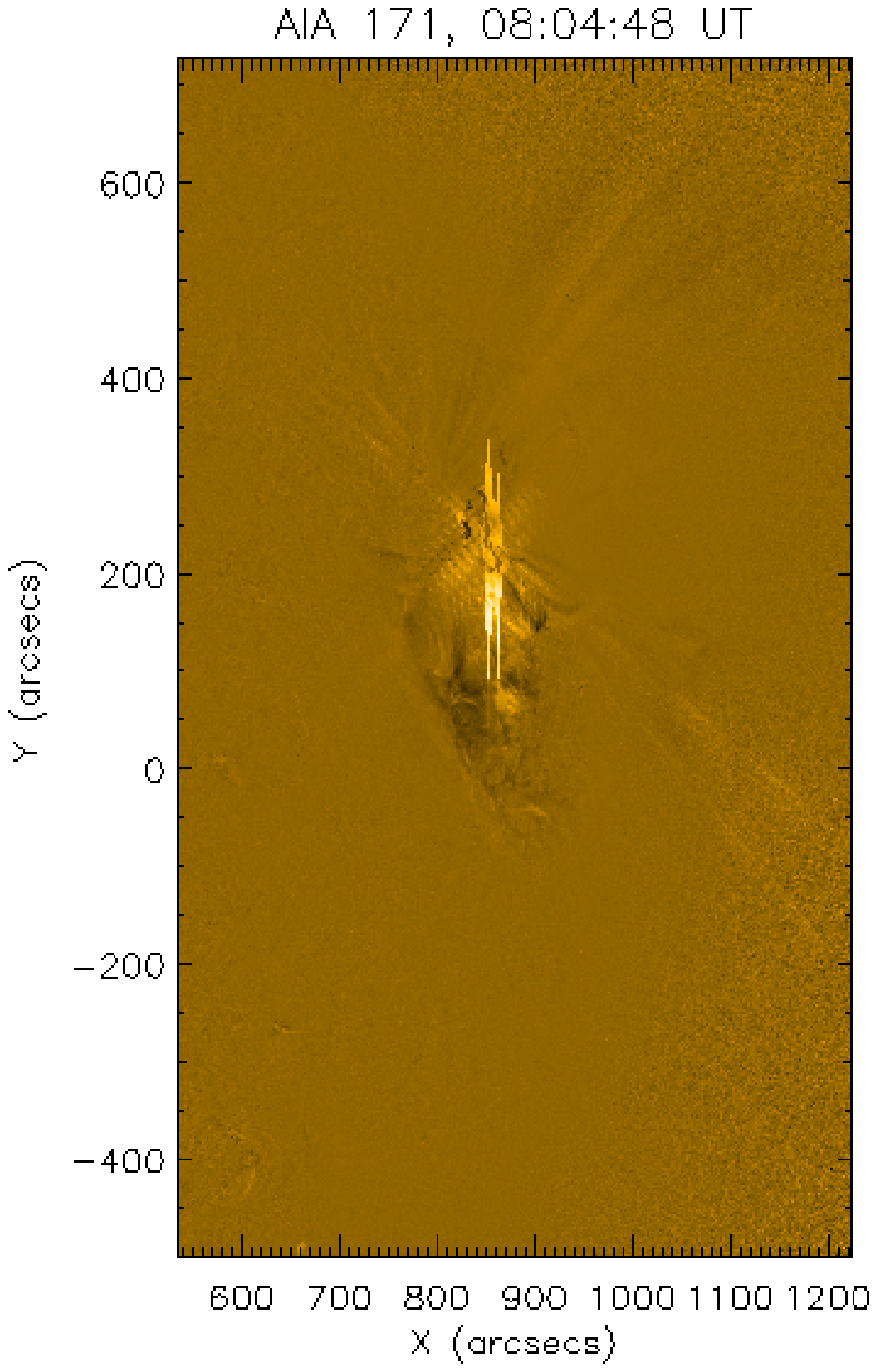}
\includegraphics[scale=0.50, angle=0]{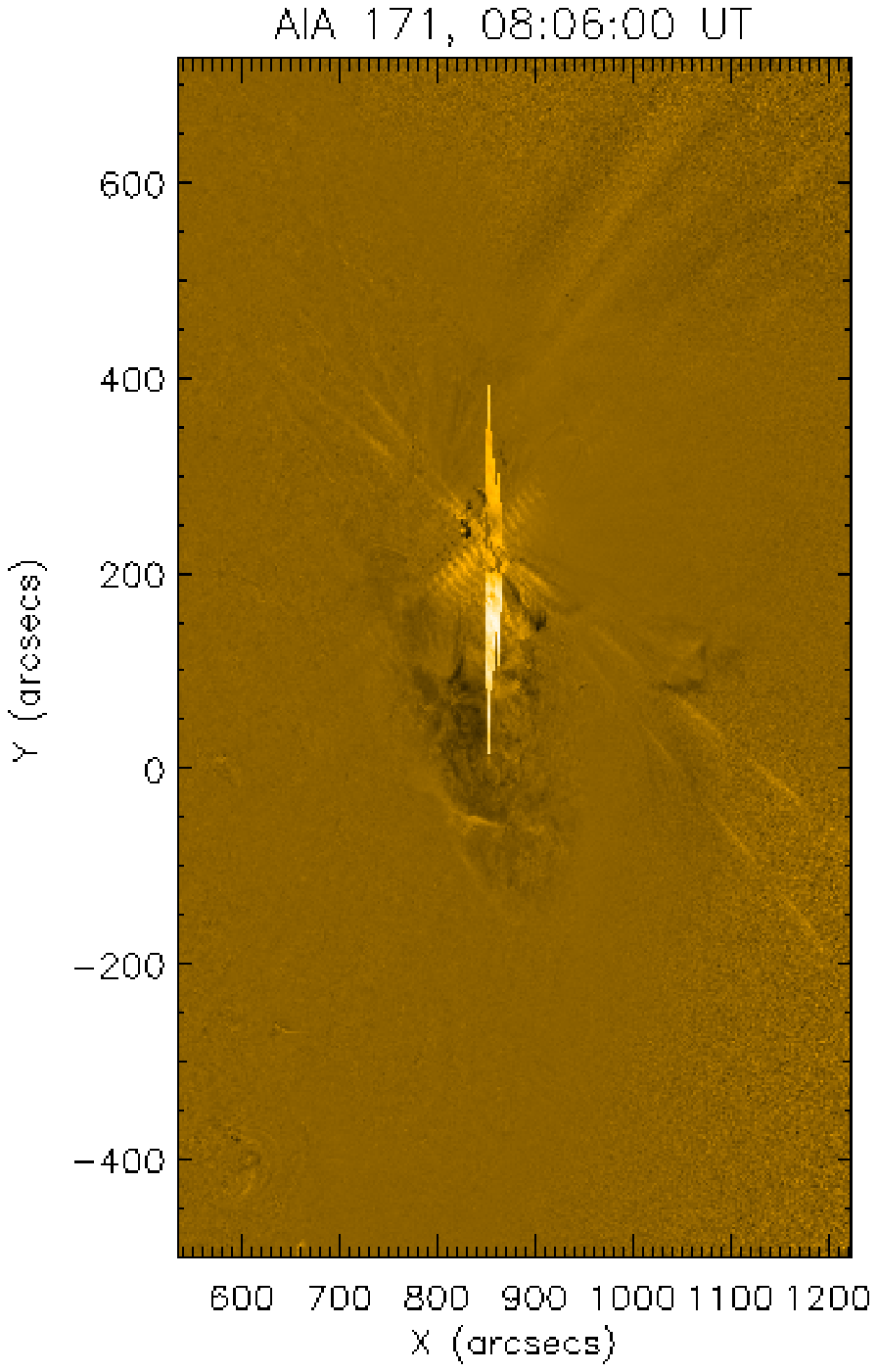}
\includegraphics[scale=0.50, angle=0]{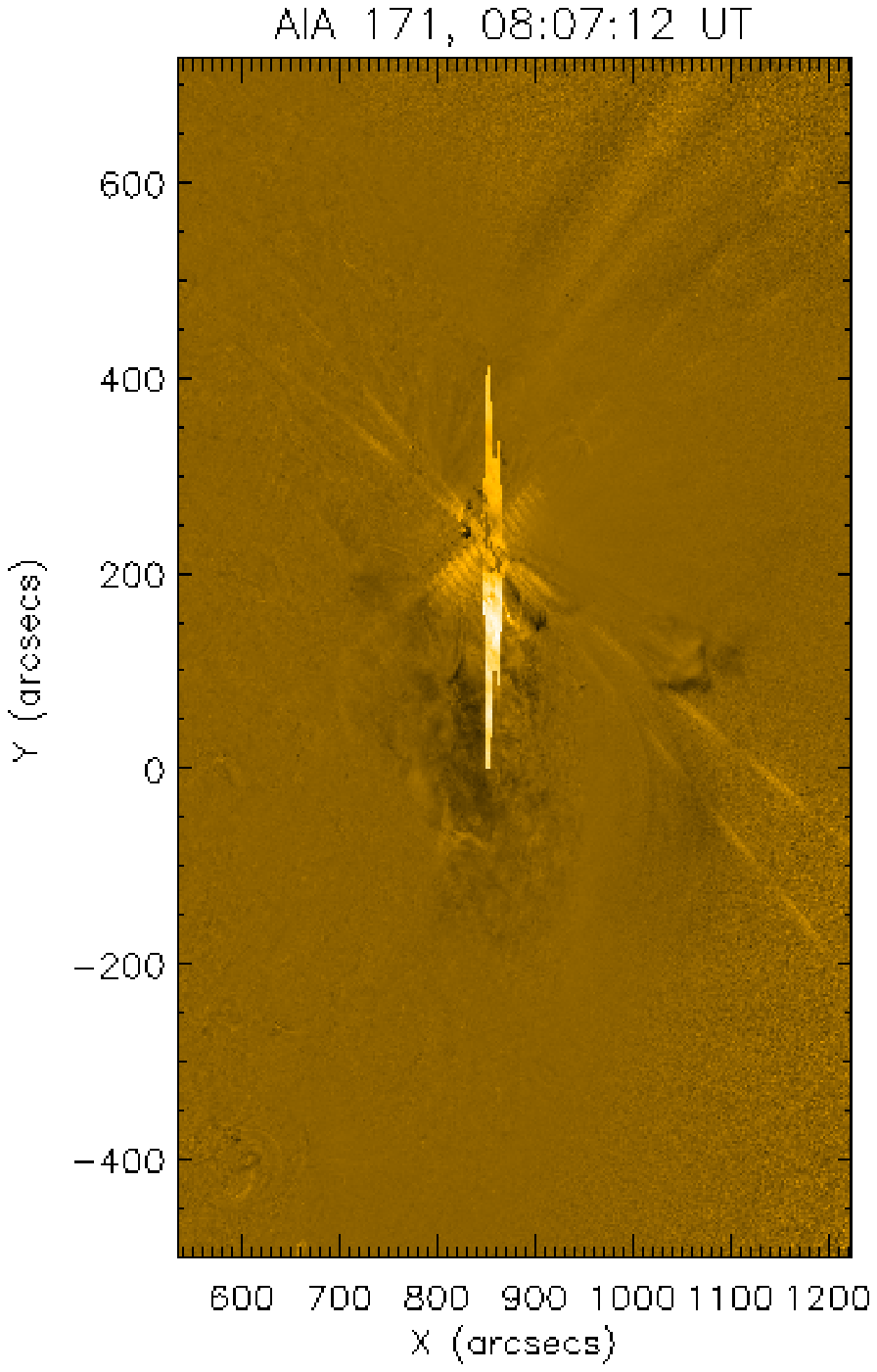}}

\centering
\mbox{
\includegraphics[scale=0.50, angle=0]{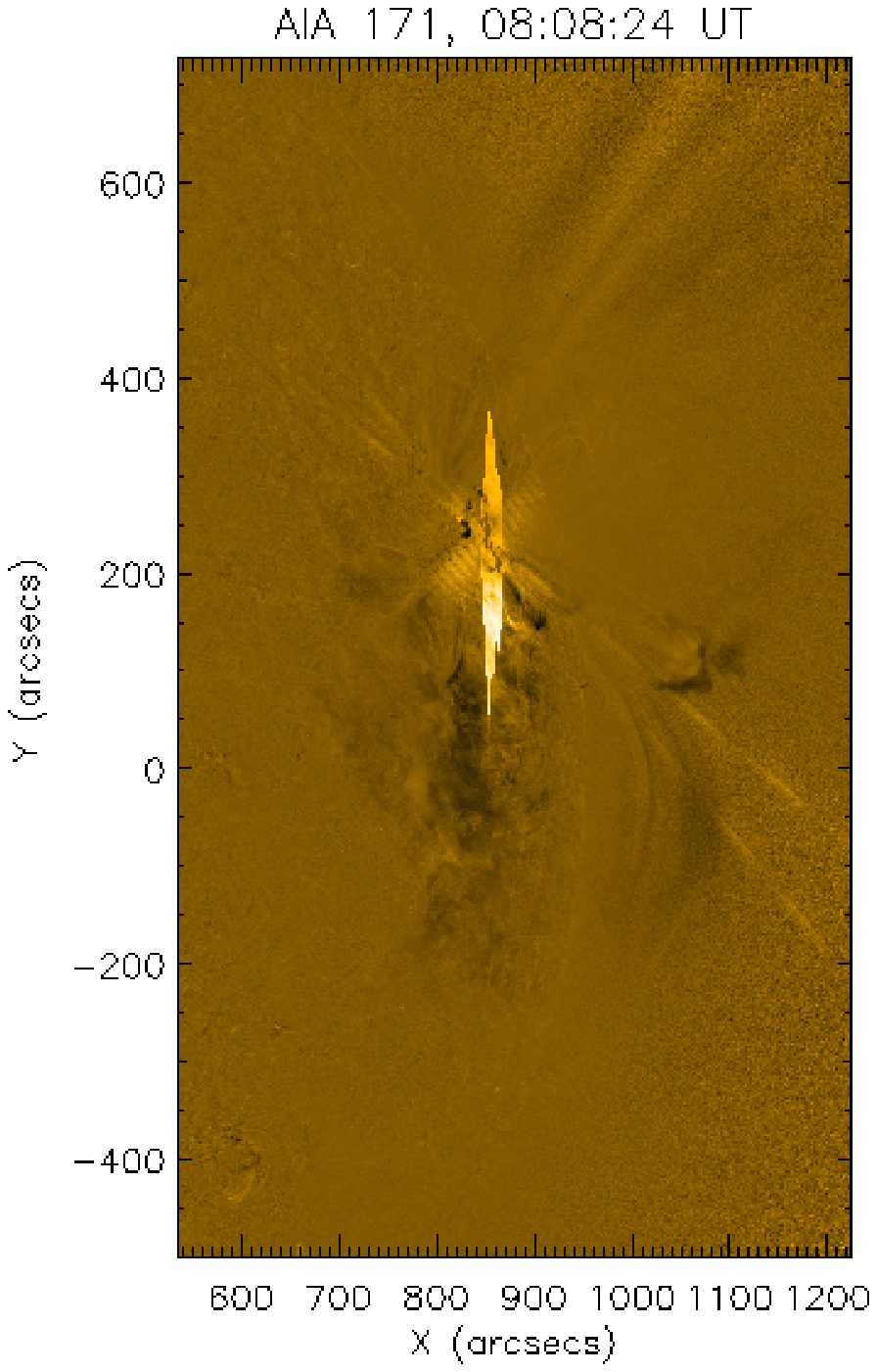}
\includegraphics[scale=0.60, angle=0]{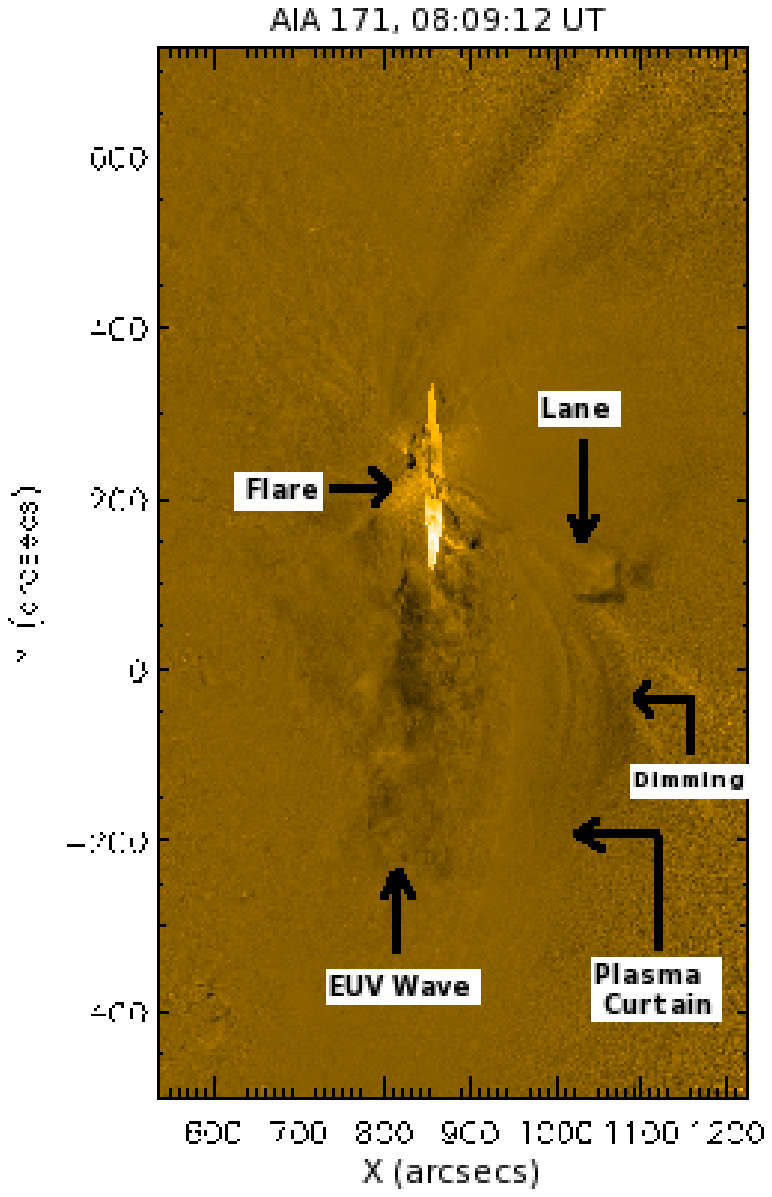}
\includegraphics[scale=0.50, angle=0]{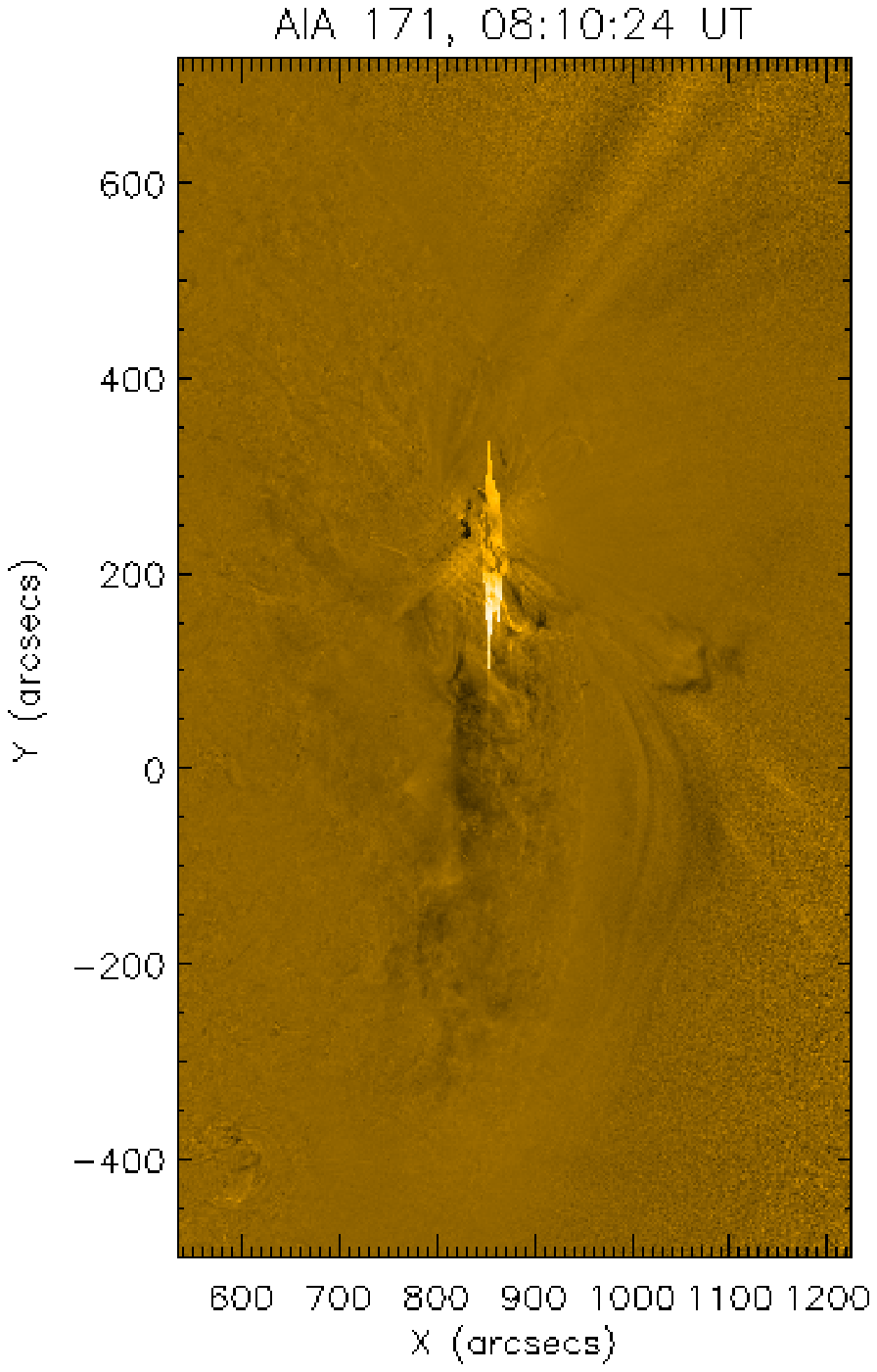}}

\caption{\small
SDO/AIA 171 \AA~difference image sequence showing the generation 
of the global coronal wave disturbances. The most of the disturbances propagate symmetrically in 
the southward direction on the disk. However, a narrow lane is also opened above the 
plasma curtain through which the part of disturbance interacts
with it and produces the transverse waves.  
}
\label{fig:JET-PULSE_1}
\end{figure*}

\clearpage

\begin{figure*}
\centering
\mbox{
\hspace{-2.0cm}
\includegraphics[scale=0.50, angle=0]{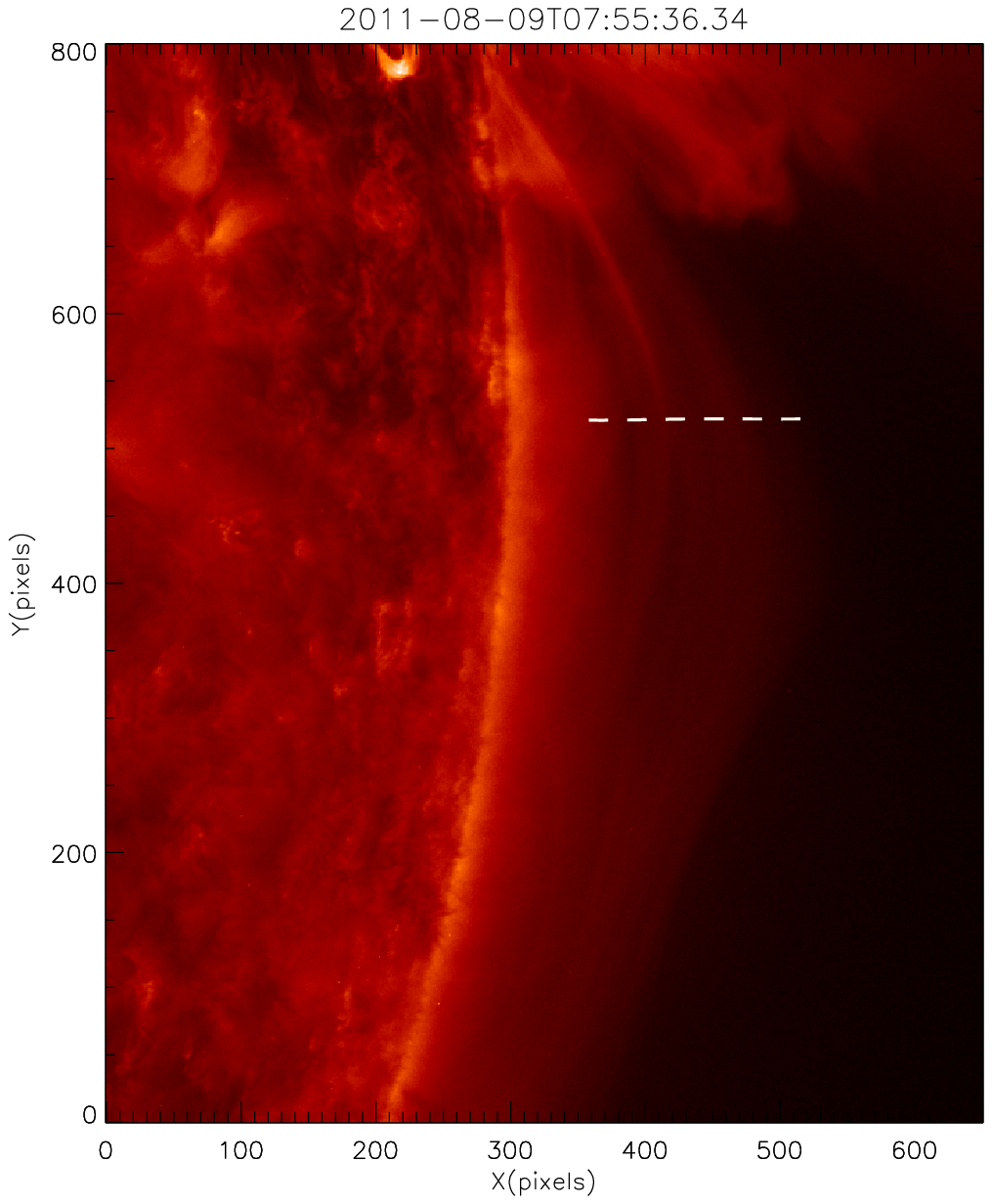}
\hspace{-4.0cm}
\includegraphics[scale=0.50, angle=0]{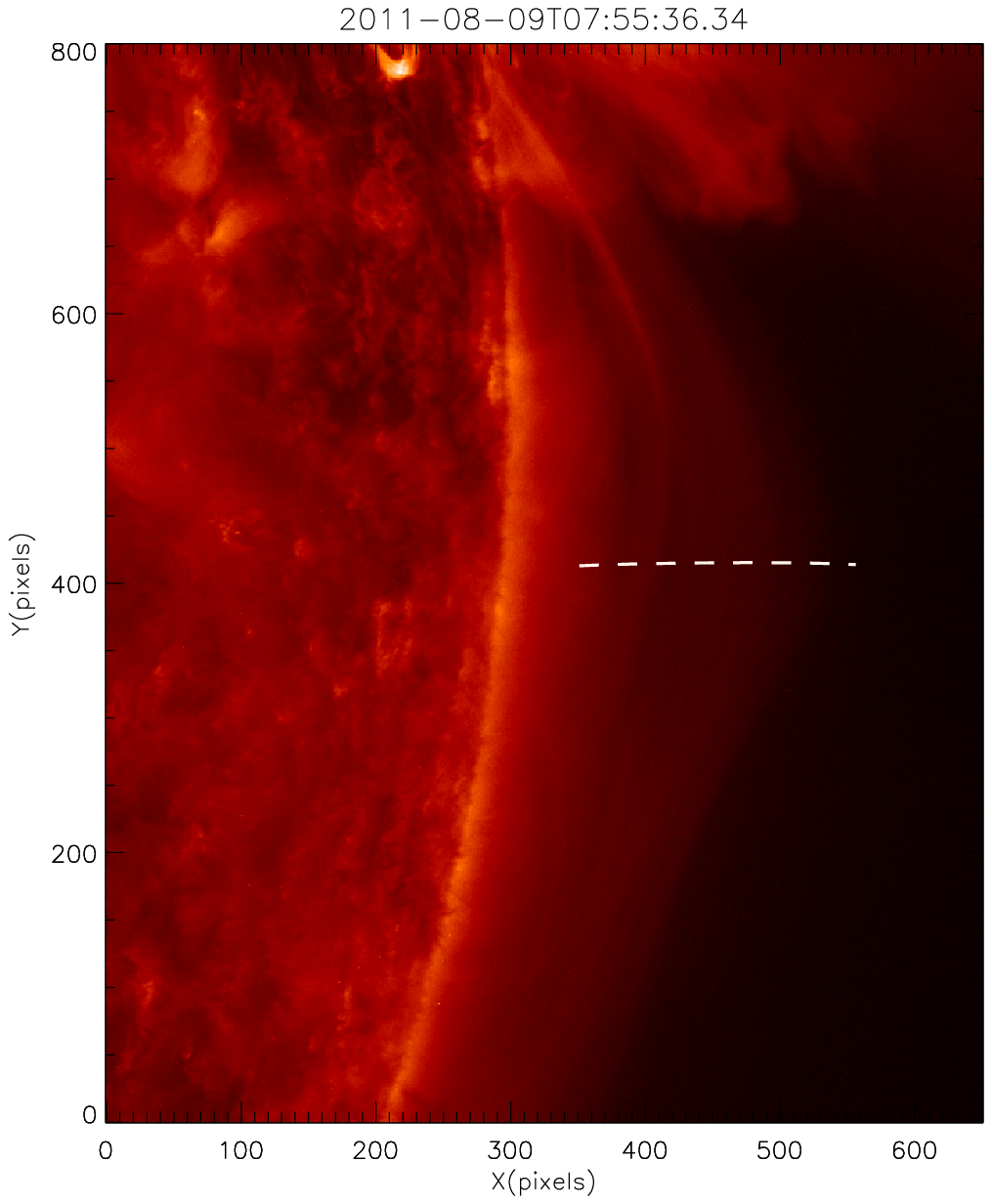}
\hspace{-2.5cm}
\includegraphics[scale=0.50, angle=0]{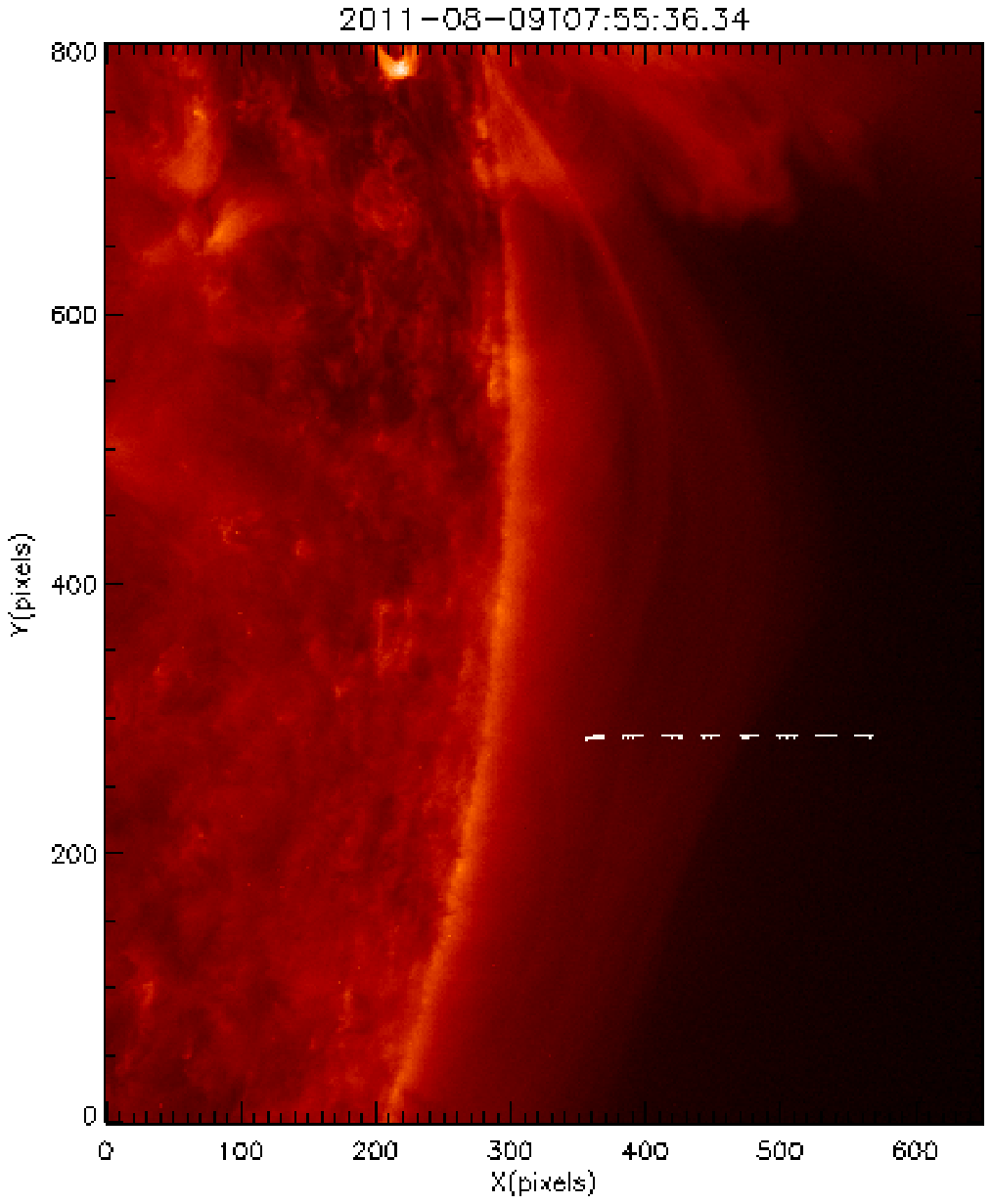}}

\centering
\mbox{
\hspace{-3.15cm}
\includegraphics[scale=0.42, angle=0]{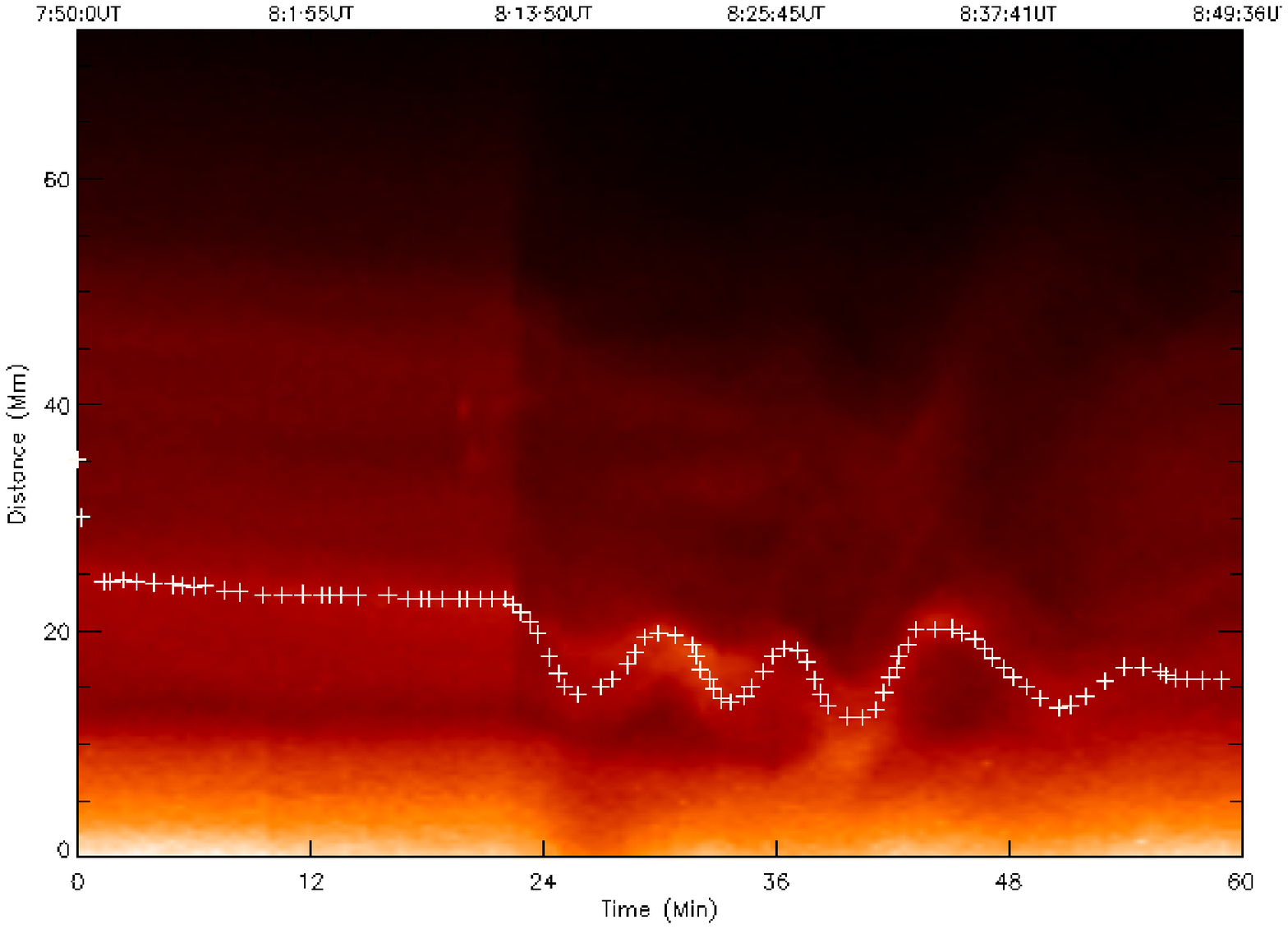}
\hspace{-0.85cm}
\includegraphics[scale=0.42, angle=0]{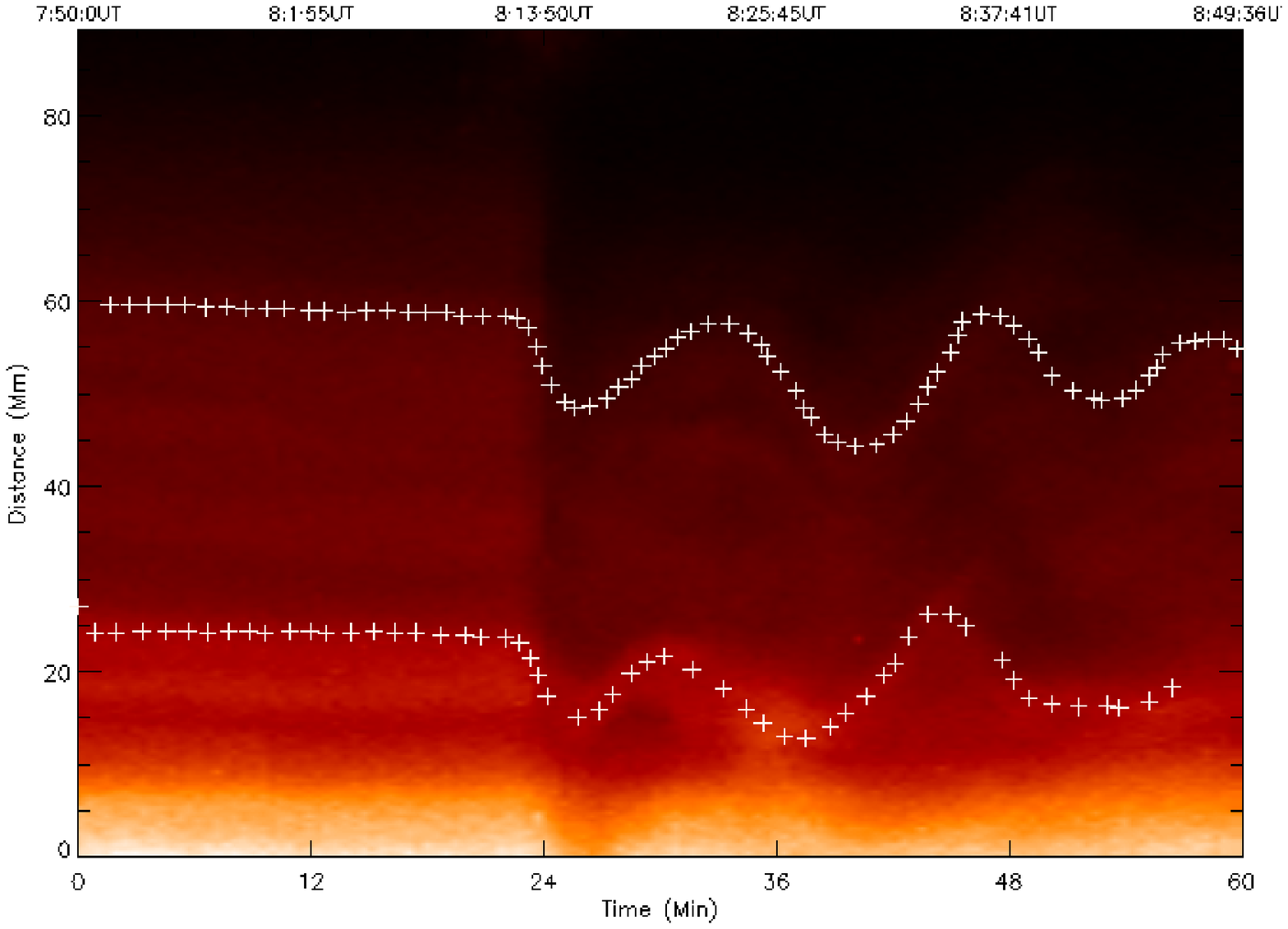}
\hspace{-0.79cm}
\includegraphics[scale=0.42, angle=0]{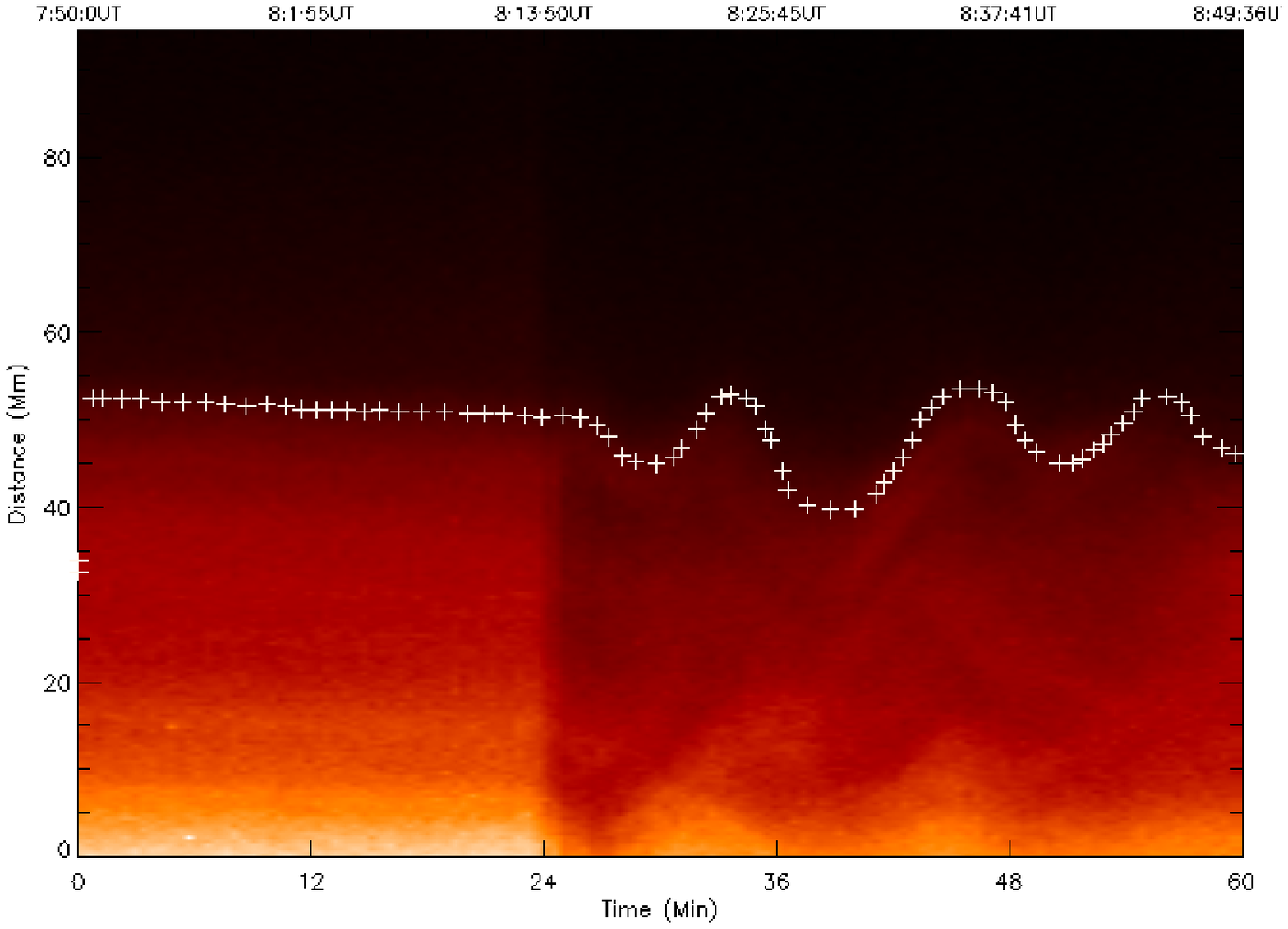}}
\caption{\small
Top three panels show the position of vertical slits over the plasma 
curtain, while bottom panels show their corressponding distance-time
diagrams. From left-to-right, the slit positions are shfting away 
from flare energy release site.
}
\label{fig:JET-PULSE_1}
\end{figure*}

\clearpage

\begin{figure*}
\centering
\mbox{
\hspace{-2.0cm}
\includegraphics[scale=0.40, angle=90]{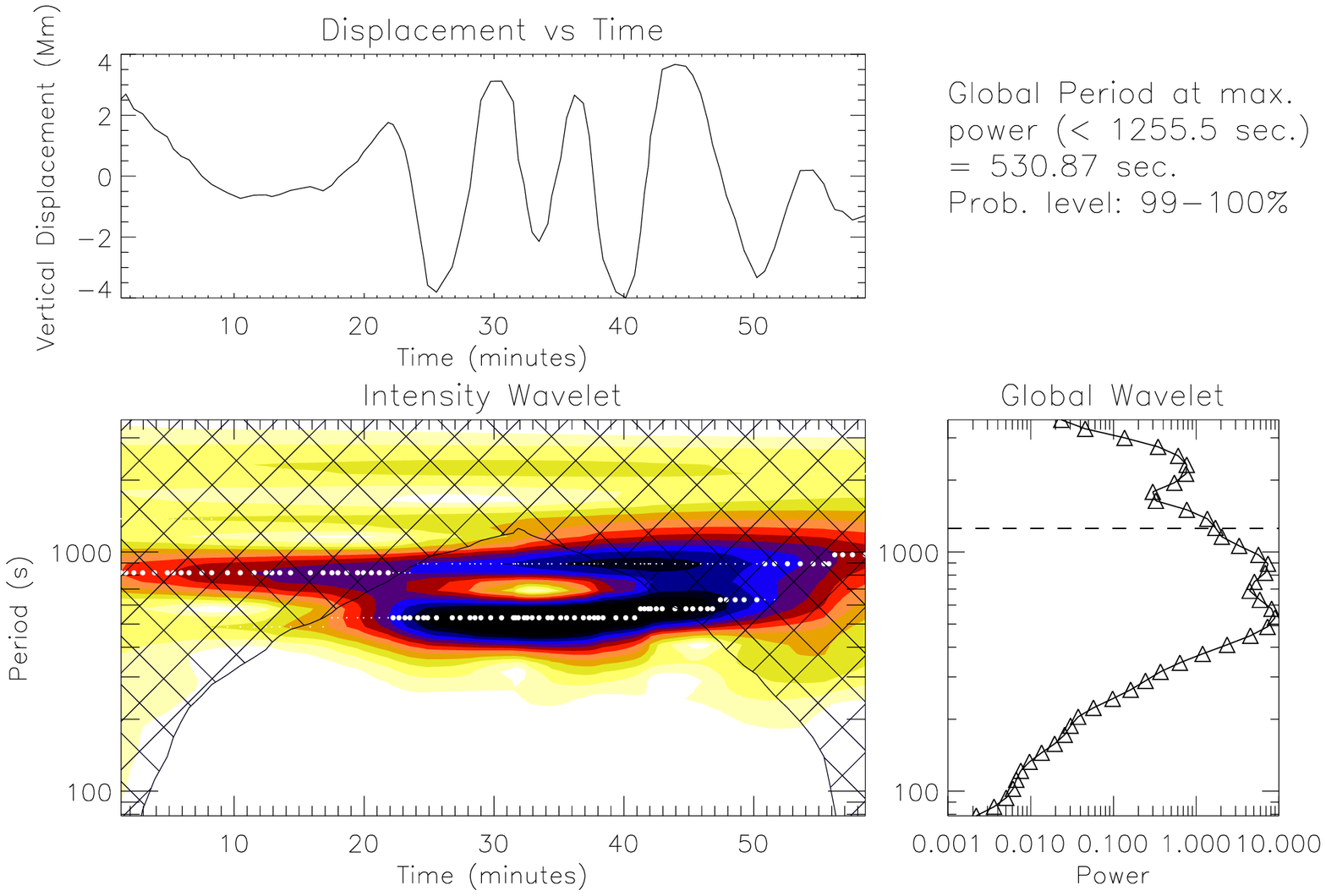}
\hspace{-2.0cm}
\includegraphics[scale=0.40, angle=90]{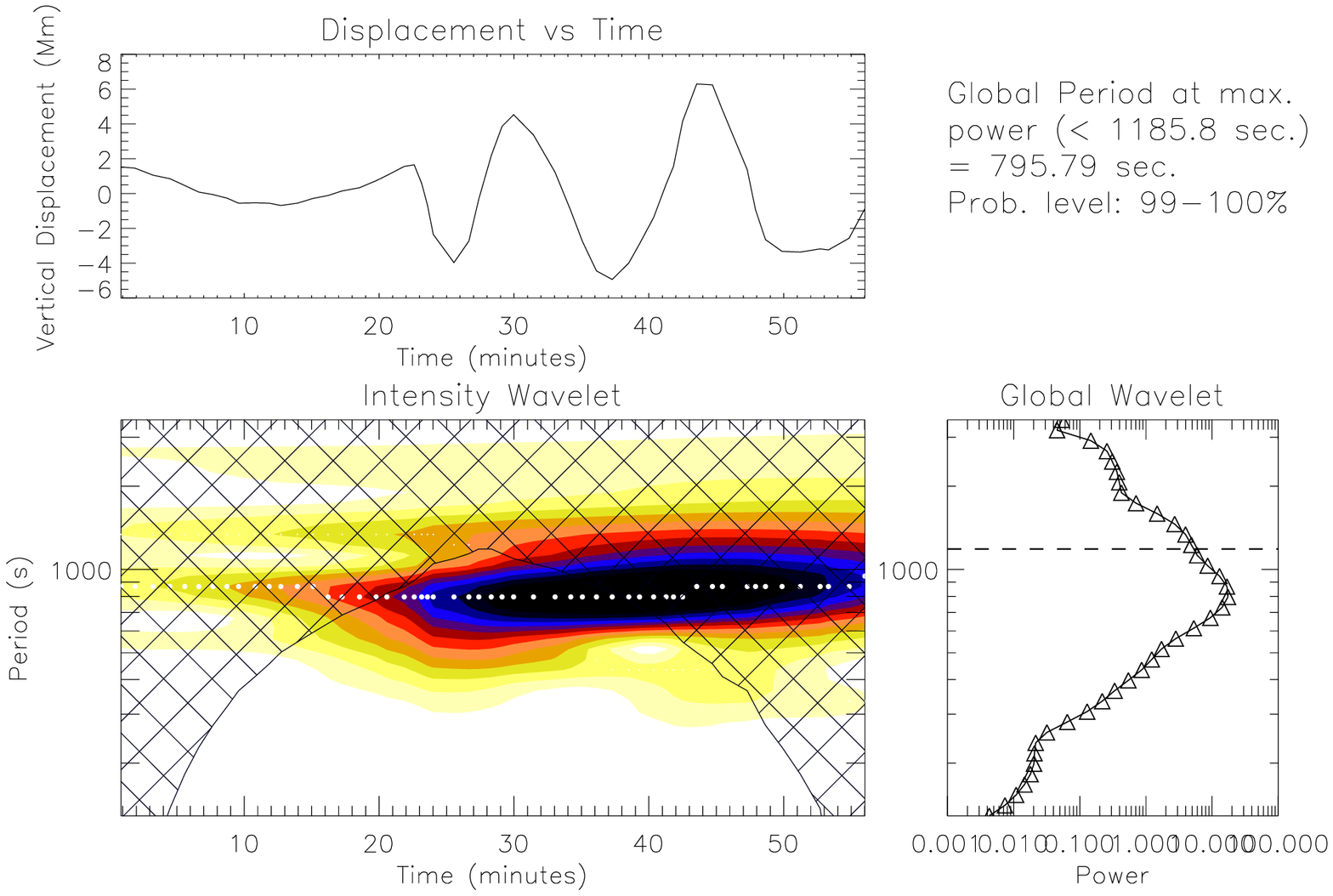}}

\caption{\small
The power spectral analyses of the transverse oscillations of deep layer of plasma curtain.
}
\label{fig:JET-PULSE_1}
\end{figure*}

\clearpage

\begin{figure*}
\centering

\mbox{
\hspace{-2.0cm}
\includegraphics[scale=0.40, angle=90]{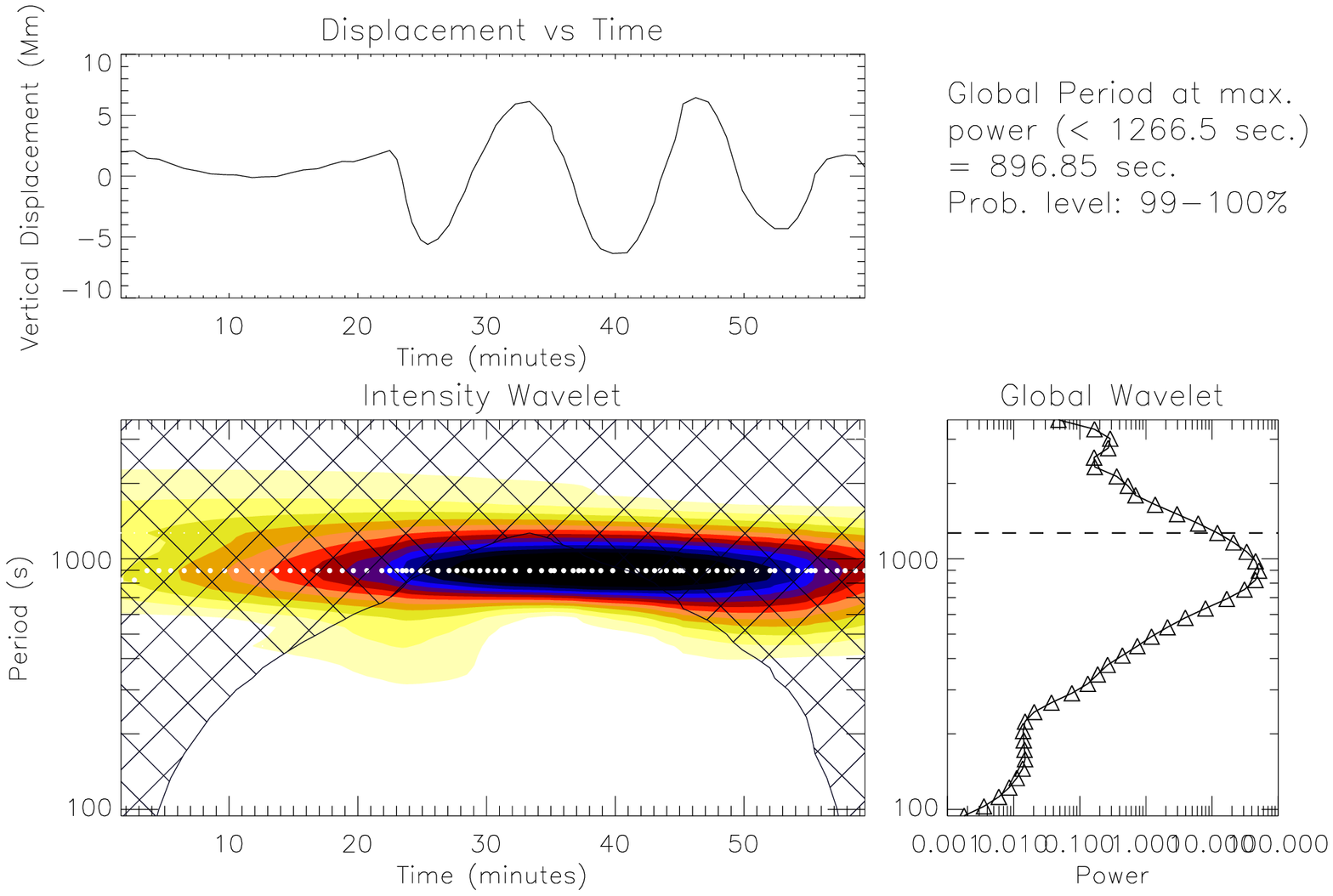}
\hspace{-2.0cm}
\includegraphics[scale=0.40, angle=90]{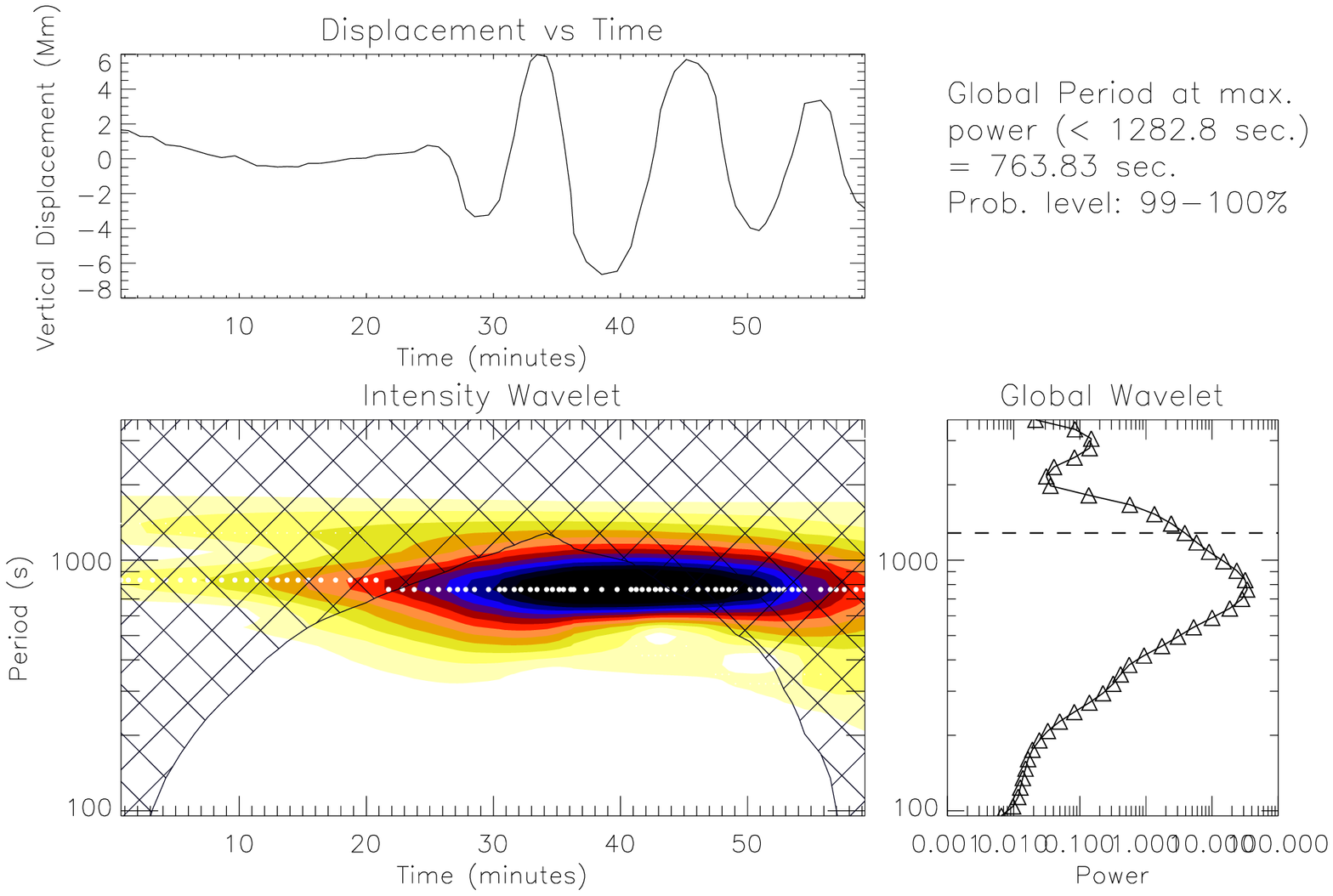}}
\caption{\small
The power spectral analyses of the transverse oscillations of the 
surface of plasma curtain at middle (left) and on its southward direction (right).
}
\label{fig:JET-PULSE_1}
\end{figure*}

\end{document}